\documentclass[sigconf, nonacm]{acmart}

\renewcommand\footnotetextcopyrightpermission[1]{}

\usepackage{amsmath, amsthm, amssymb, amsfonts} \usepackage{algpseudocode} \usepackage{algorithm} \usepackage{graphicx} \usepackage{multirow} \usepackage{enumitem}
\usepackage{setspace}

\AtBeginDocument{%
	\providecommand\BibTeX{{%
			\normalfont B\kern-0.5em{\scshape i\kern-0.25em b}\kern-0.8em\TeX}}}

\begin{document}
\title{Hyperbolic Neural Collaborative Recommender}

\author{Anchen Li}
\affiliation{%
	\institution{College of Computer Science and Technology, Jilin University, China}}
\email{liac20@mails.jlu.edu.cn}

\author{Bo Yang}
\authornote{Corresponding author.}
\affiliation{%
	\institution{College of Computer Science and Technology, Jilin University, China}}
\email{ybo@jlu.edu.cn}

\author{Hongxu Chen}
\affiliation{%
	\institution{University of Technology Sydney, Australia}}
\email{Hongxu.Chen@uts.edu.au}

\author{Guandong Xu}
\affiliation{%
	\institution{University of Technology Sydney, Australia}}
\email{Guandong.Xu@uts.edu.au}


\begin{abstract}
This paper explores the use of hyperbolic geometry and deep learning techniques for recommendation. We present \textbf{H}yperbolic \textbf{N}eural \textbf{C}ollaborative \textbf{R}ecommender (HNCR), a deep hyperbolic representation learning method that exploits mutual semantic relations among users/items for collaborative filtering (CF) tasks. HNCR contains two major phases: neighbor construction and recommendation framework. The first phase introduces a neighbor construction strategy to construct a semantic neighbor set for each user and item according to the user-item historical interaction. In the second phase, we develop a deep framework based on hyperbolic geometry to integrate constructed neighbor sets into recommendation. Via a series of extensive experiments, we show that HNCR outperforms its Euclidean counterpart and state-of-the-art baselines.

\end{abstract}

\maketitle
\section{Introduction}
In the era of information explosion, recommender systems have been playing an indispensable role in meeting user preferences by recommending products or services. Collaborative filtering (CF), which focuses on utilizing the historical user-item interactions to generate recommendations, remains to be a fundamental task towards effective personalized recommendation \cite{CF1,CF2,CF3,CF4}. 

User and item embedding learning is the key to CF. Early models like Matrix Factorization (MF) embed users and items in a shared latent space and model the user preference to an item as the inner product between user and item embeddings \cite{MF}. However, due to the complex interaction between users and items, the shallow representations in MF-based methods lack expressiveness to model features \cite{NCF,MMCF}. As deep learning developed, some recommendation approaches utilize neural networks to capture complex interaction behaviors, which enhance the performance of previous shallow models \cite{DMF,NFM,NCF,NGCF,LRGCCF,CMN,MMCF}.

Notably, most of the existing deep CF models primarily operate in Euclidean spaces. From the perspective of the graph, the user-item interactions can be considered as a bipartite graph, which in turn renders a so-called complex network \cite{BipaUI}. The properties of complex networks have been widely studied before, and it is known that they are closely related to hyperbolic geometry \cite{hp1}. Moreover, real-world user-item interaction relations often exhibit the power-law distribution. Recent research shows that hyperbolic geometry enables embeddings with much smaller distortion when embedding data with the power-law distribution \cite{Poincare,HGCN}. This motivates us to consider whether we can combine hyperbolic geometry and deep learning techniques for boosting performance.

\par
\begin{figure}[h!]
	\centering
	\includegraphics[width=0.98\linewidth]{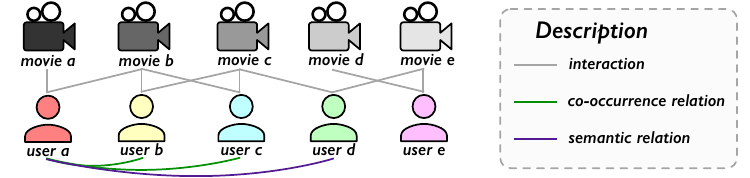}
	\caption{A simple user-item interaction scenario.}
	\label{fig:example}
\end{figure}

   
In addition, most deep recommendation models have less focus on explicitly modeling user-user or item-item high-order semantic correlations, while such relations could provide valuable information to inference user or item features. Although some existing works \cite{CMN,MMCF} utilize the co-occurrence relation (co-engage between users or co-engaged between items) to define the neighbors for users and items, we argue that such co-occurrence relation is macro-level and coarse-grained. For instance, Figure~\ref{fig:example}(a) shows a simple user-item interaction in the movie domain. Although user $a$ and user $d$ are not co-occurrence relations, they both share common preferences with user $b$ and user $c$. Such high-order semantic correlation is also a very significant signal for revealing user preferences and item properties, while it is ignored by most existing works. Moreover, for inactive users and items, their co-occurrence relations may be sparse, which is insufficient to provide complementary information for these users and items.

Motivated by the above observations, we propose to consider user-user and item-item semantic correlations and utilize hyperbolic representation learning to integrate them into recommendation. Indeed, it is a non-trivial task due to the following two key challenges. First, as hyperbolic spaces are not vector spaces, Euclidean embedding operations (e.g., addition, multiplication) for users and items cannot be carried in hyperbolic spaces. How can we effectively design a deep recommendation model in hyperbolic spaces in an elegant way? Second, mutual semantic relations among users/items are latent in user-item interactions and difficult to capture directly. A recent work \cite{Gemo} proposes to map the original graph to a latent space and then find nodes' semantic neighbors in the latent space, while their methods cannot be directly applied to our setting, as there is no explicit user/item graph. Thus, how to build user/item semantic neighborhoods becomes another challenge.

\textit{\textbf{Our approach.}} To solve the above technical challenges, in this paper, we present a novel method \textbf{H}yperbolic \textbf{N}eural \textbf{C}ollaborative \textbf{R}ecommender (HNCR) for the CF task. Our method HNCR contains two phases. In the first phase, HNCR devises a neighbor construction strategy to find semantic neighbors for users and items. More specifically, HNCR first utilizes the user-item interaction information to construct a user relational graph and an item relational graph and then map each graph to a latent continuous space to find neighbors with semantic transitive relation for users and items. In the second phase, HNCR carries a deep framework based on hyperbolic representation learning to integrate constructed neighbor sets into recommendation. To support vector operations, we utilize the operations of gyrovector spaces \cite{gyrovector1,gyrovector2} to build our framework. 

\textit{\textbf{Our contributions.}} In summary, our main contributions in this paper are listed as follows:
\begin{itemize}[leftmargin= 8 pt]
\item We propose a hyperbolic neural approach HNCR that explicitly models mutual semantic relations among users/items for CF tasks.
\item We introduce a method to find user and item semantic neighbors.
\item We propose a framework based on hyperbolic geometry, which employs gyrovector space operations to integrate constructed semantic neighbors into recommendations.
\item Experimental results on four datasets show HNCR not only outperforms its Euclidean counterpart but also boosts the performance over the state-of-the-art approaches. 
\end{itemize}

\section{Background}
\label{sec:background}
In this section, we review the background of hyperbolic geometry and gyrovector space, which forms the basis of our method.

\subsection{Hyperbolic Geometry}
The hyperbolic space is uniquely defined as a complete and simply connected Riemannian manifold with constant negative curvature \cite{hp1}. A key property of hyperbolic spaces is that they expand faster than Euclidean spaces. To describe the hyperbolic space, there are multiple commonly used models of hyperbolic geometry, such as the Poincar$\acute{\text{e}}$ model, hyperboloid model, and Klein model \cite{hp2}. These models are all connected and can be converted into each other. In this paper, we work with the Poincar$\acute{\text{e}}$ ball model because it is well-suited for gradient-based optimization \cite{Poincare}. 

\textit{Poincar$\acute{\text{e}}$ ball model.}
Let $\mathbb{D}^n = \left\{  \textbf{x} \in \mathbb{R}^n: \left \| \textbf{x} \right \| < 1  \right\}$ be the $open$ $n$-dimensional unit ball, where $\left \| \cdot \right \|$ denotes the Euclidean norm. The Poincar$\acute{\text{e}}$ ball model is the Riemannian manifold $(\mathbb{D}^n, g^{\mathbb{D}})$, which is defined by the manifold $\mathbb{D}^n$ equipped with the Riemannian metric tensor $g_{\textbf{x}}^{\mathbb{D}} = \lambda_\textbf{x}^2g^{\mathbb{E}}$, where $\lambda_\textbf{x} = \frac{2}{1-\left \| \textbf{x} \right \|^2}$; $\textbf{x} \in \mathbb{D}^n$; and $g^{\mathbb{E}} = \textbf{I}$ denotes the Euclidean metric tensor.

\subsection{Gyrovector Spaces}
The framework of gyrovector spaces provides vector operations for hyperbolic geometry \cite{HNN}. We will make extensive use of these gyrovector operations to design our model. Specifically, these operations in gyrovector spaces are defined in an open $n$-dimensional ball $\mathbb{D}^n_c = \left\{  \textbf{x} \in \mathbb{R}^n: c \left \| \textbf{x} \right \|^2 < 1  \right\}$ of radius $\frac{1}{\sqrt{c}}(c \geq 0 )$. Some widely used vector operations of gyrovector spaces are defined as follows:
\begin{itemize}[leftmargin= 12 pt]
\item \textit{M{\"o}bius addition}: For $\textbf{x}, \textbf{y} \in \mathbb{D}^n_c$, the M{\"o}bius addition of $\textbf{x}$ and $\textbf{y}$ is defined as follows:
\begin{align}
\textbf{x} \oplus_c \textbf{y} = \frac{(1+2c\left \langle \textbf{x},\textbf{y} \right \rangle + c\left \| \textbf{y} \right \|^2)\textbf{x} + (1-c\left \| \textbf{x} \right \|^2)\textbf{y}}{1+2c\left \langle \textbf{x},\textbf{y} \right \rangle + c^2\left \| \textbf{x} \right \|^2 \left \| \textbf{y} \right \|^2}.
\end{align}
In general, this operation is not commutative nor associative. 
\item \textit{M{\"o}bius scalar multiplication}: For $c > 0$, the M{\"o}bius scalar multiplication of $\textbf{x} \in \mathbb{D}^n_c \backslash \left \{ \textbf{0} \right \}$ by $r \in \mathbb{R}$ is defined as follows:
\begin{align}
r \otimes_c \textbf{x} = \frac{1}{\sqrt{c}}\tanh\left ( r \tanh^{-1} \left ( \sqrt{c} \left \| \textbf{x} \right \| \right ) \right ) \frac{\textbf{x}}{\left \| \textbf{x} \right \|},
\end{align}
and $r \otimes_c \textbf{0} = \textbf{0}$. This operation satisfies associativity: 
\item \textit{M{\"o}bius matrix-vector multiplication}: For $\textbf{M} \in \mathbb{R}^{n' \times n}$ and $\textbf{x} \in \mathbb{D}^n_c$, if $\textbf{Mx} \neq \textbf{0}$, the M{\"o}bius matrix-vector multiplication of $\textbf{M}$ and $\textbf{x}$ is defined as follows:
\begin{align}
\textbf{M} \otimes_c \textbf{x} = \frac{1}{\sqrt{c}}\tanh\left ( \frac{\left \| \textbf{Mx} \right \|}{\left \| \textbf{x} \right \|} \tanh^{-1} \left ( \sqrt{c} \left \| \textbf{x} \right \| \right ) \right ) \frac{\textbf{Mx}}{\left \| \textbf{Mx} \right \|}.
\end{align}
This operation satisfies associativity.
\item \textit{M{\"o}bius exponential map and logarithmic map}: For $\textbf{x} \in \mathbb{D}^n_c$, it has a tangent space $T_\textbf{x}\mathbb{D}^n_c$ which is a local first-order approximation of the manifold $\mathbb{D}^n_c$ around $\textbf{x}$. The logarithmic map and the exponential map can move the representation between the two manifolds in a correct manner. For any $\textbf{x} \in \mathbb{D}^n_c$, given $\textbf{v} \neq \textbf{0}$ and $\textbf{y} \neq \textbf{x}$, the M{\"o}bius exponential map $\exp_{\textbf{x}}^c: T_\textbf{x}\mathbb{D}^n_c \rightarrow \mathbb{D}^n_c$ and logarithmic map $\log_{\textbf{x}}^c: \mathbb{D}^n_c \rightarrow T_\textbf{x}\mathbb{D}^n_c$ are defined as follows:
\begin{align} 
\exp_{\textbf{x}}^c(\textbf{v}) = \textbf{x} \oplus_c \left(\tanh\left(\sqrt{c}\frac{\lambda_\textbf{x}^c\left \| \textbf{v} \right \|}{2}\right)\frac{\textbf{v}}{\sqrt{c}\left \| \textbf{v} \right \|}\right),
\end{align}
\begin{align}
\log_{\textbf{x}}^c(\textbf{y}) = \frac{2}{\sqrt{c}\lambda_\textbf{x}^c} \tanh^{-1}(\sqrt{c}\left \| -\textbf{x} \oplus_c \textbf{y} \right \|) \frac{-\textbf{x} \oplus_c \textbf{y}}{\left \| -\textbf{x} \oplus_c \textbf{y} \right \|},
\end{align}
where $\lambda_\textbf{x}^c  = \frac{2}{1-c\left \| \textbf{x} \right \|^2}$ is the conformal factor of $(\mathbb{D}^n_c, g^c)$, where $g^c$ is the generalized hyperbolic metric tensor. 
\item \textit{Distance}: For $\textbf{x}, \textbf{y}\in \mathbb{D}^n_c$, the generalized distance between them in Gyrovector spaces are defined as follows:
\begin{align}
d_{c}(\textbf{x}, \textbf{y}) = \frac{2}{\sqrt{c}}\tanh^{-1}(\sqrt{c}\left \| -\textbf{x} \oplus_c \textbf{y} \right \|).
\end{align}
\end{itemize}

We will make use of these M{\"o}bius gyrovector space operations to design our recommendation framework.
\section{Methodology}
\label{sec:methodology}
In this section, we first introduce the notations and formulate the problems. We then describe two phases of HNCR: (i) neighbor construction and (ii) recommendation framework. 

\subsection{Notations and Problem Formulation}
In a typical recommendation scenario, we suppose there are $M$ users $\mathcal{U}$ = $\left\{ u_{1},u_{2},...,u_{M} \right\}$ and $N$ items $\mathcal{V}$ = $\left\{ v_{1},v_{2},...,v_{N} \right\}$. We define \textbf{Y} $\in \mathbb{R}^{M \times N}$ as the user-item interaction matrix whose element $y_{ai} \in \left\{0, 1\right\}$ indicates whether $u_a$ has engaged with $v_i$ or not. 

\par
Given the above information ($\mathcal{U}, \mathcal{V}, \textbf{Y}$), the first phase of HNCR outputs user relational data $\mathcal{N}_u = \left\{ \mathcal{N}_u(1),\mathcal{N}_u(2),...,\mathcal{N}_u(M) \right\}$ and item relational data $\mathcal{N}_v = \left\{ \mathcal{N}_v(1),\mathcal{N}_v(2),...,\mathcal{N}_v(N) \right\}$. $\mathcal{N}_u$ and $\mathcal{N}_v$ contain semantic neighbors for users and items, respectively. The details of building up $\mathcal{N}_u$ and $\mathcal{N}_v$ are discussed in Section~\ref{sec:neighbor_construction}. 

\par 
In the HNCR's second phase, given the interaction matrix \textbf{Y}, user neighbor data $\mathcal{N}_u$ and item neighbor data $\mathcal{N}_v$, the recommendation framework aims to learn a prediction function $\hat{y}_{ai} = \mathcal{F}(u_a, v_i | \Theta, \textbf{Y}, \mathcal{N}_u, \mathcal{N}_v)$, where $\hat{y}_{ai}$ is the preference probability that $u_a$ will engage with $v_i$, and $\Theta$ is the framework parameters of the function $\mathcal{F}$. The details of this phase are discussed in Section~\ref{sec:framework}.

\subsection{Neighbor Construction}
\label{sec:neighbor_construction}
\begin{figure}[h!]
	\centering
	\includegraphics[width=0.99\linewidth]{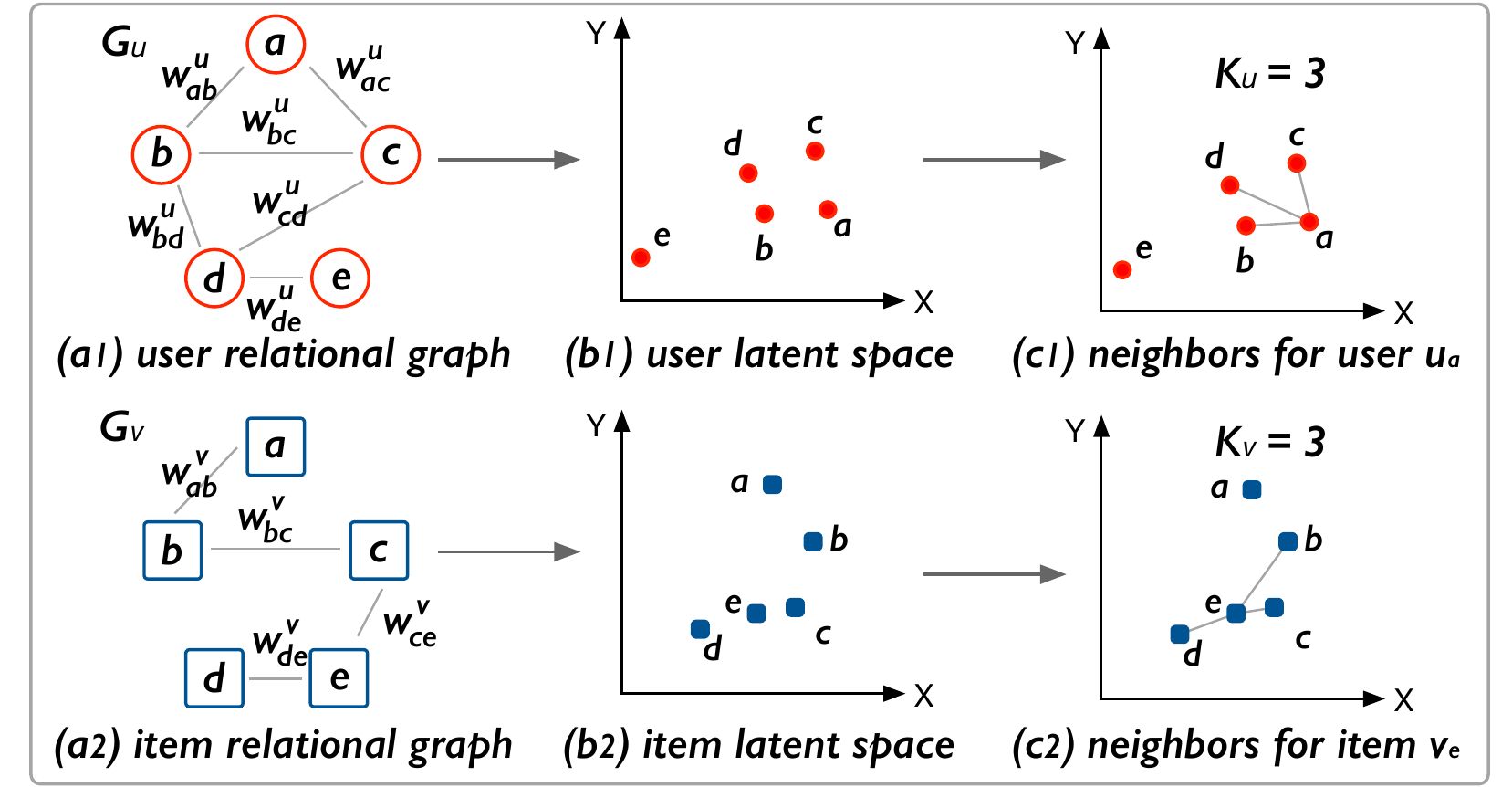}
	\caption{An illustration of neighbor construction. (a) $\to$ (b): map user and item relational graphs to latent spaces respectively. (b) $\to$ (c): find semantic neighbors for users and items. }
	\label{fig:neighbor}
\end{figure}

In this subsection, we describe the neighbor construction strategy of HNCR. This strategy contains three major steps: (i) construct the user relational graph and item relational graph; (ii) map user and item relational graphs to latent spaces respectively; and (iii) find semantic neighbors for users and items from their latent spaces. In this work, the construction of the relational data is constrained to utilizing the user-item interaction records in the training split. Figure~\ref{fig:neighbor} illustrates the process of neighbor construction strategy. In the following statement, we illustrate the process for the user side and the same process works for the item side.

\subsubsection{Step 1: Construction of relational graphs}
The user-item interaction data can be represented by a bipartite graph structure. We first transform the bipartite graph to construct the user relational graph $\mathcal{G}_u = (\mathcal{U}, \mathcal{E}_u)$ for identifying user-user relationships. In the user relational graph $\mathcal{G}_u$, the edge $e^u \in \mathcal{E}_u$ connects two users if they at least engaged one common item before. In addition, $e^u$ is associated with a weight $w^u > 0$ to indicate the relational strength between two users. We design a delicate method to define the weight, which can represent the relations between users in a fine-grained way. Specifically, we define the weight $w^u_{ab}$ for the edge $e^u_{_{ab}}$ between $u_a$ and $u_b$ as $w^u_{ab} = h^u_{ab} \cdot c^u_{ab}$, which is determined by two aspects, the historical behavior factor $h^u_{ab}$ and the popularity of co-interacted items factor $c^u_{ab}$. 

For the first aspect, if the historical behaviors of $u_a$ and $u_b$ are similar, the weight $w^u_{ab}$ should be large, and vice versa. We use heat kernel to define $h^u_{ab}$ as follows:
\begin{align}
h^u_{ab} = e^{-\frac{\left \| \textbf{Y}_a - \textbf{Y}_b \right \|^2}{t}},
\end{align}
where $\textbf{Y}_a$ and $\textbf{Y}_b$ are the corresponding rows in the user-item interaction matrix $\textbf{Y}$ for $u_a$ and $u_b$. $t$ is the time parameter in the heat conduction equation and we set $t = 100$. 

For the second aspect, if the co-interacted items are unpopular, the weight $w^u_{ab}$ should be large, and vice versa. This is because unpopular items can better reflect users' personalized preferences \cite{LightGCN}. We define $c^u_{ab}$ as follows:
\begin{align}
c^u_{ab} = \frac{2}{|\mathcal{C}_{ab}|}\cdot\sum\nolimits_{v_i\in\mathcal{C}_{ab}}\frac{1}{|\mathcal{I}_v(i)|},
\end{align}
where $\mathcal{C}_{ab}$ contains items that rated by $u_a$ and $u_b$ and $\mathcal{I}_v(i)$ contains users that rated $v_i$ before.

Figure~\ref{fig:neighbor}(a1) and (a2) are toy examples of building relational graphs for users and items in the user-item interaction scenario of Figure~\ref{fig:example}. The advantage is that such relational graphs not only reflect the strength of co-occurrence relations (one-hop neighbors) but also infers high-order semantic relations (multi-hop neighbors).

\subsubsection{Step 2: Relational graph mapping}
After the relational graph construction, we utilize the node embedding method to map the relational graph to a latent continuous space. Specifically, for the user relational graph, we use a function $f_u: u\rightarrow z^u$ to map a user node $u \in \mathcal{U}$ from $\mathcal{G}_u$ to a low-dimensional vector $z^u \in \mathbb{R}^{l_u}$ in a latent continuous space, where $l_u$ is the dimension number of the vector for users. After the mapping, both structures and properties of relational graphs are preserved and presented as the geometry in the latent space. Also, for the target user, users with important high-order transitive semantic relations will appear near the target user, while nodes with irrelevant information will appear far away from the target users. Recent research reveals that a common embedding method which only preserves the connection patterns of a graph can be effective \cite{Gemo}. Since node embedding is not the main concern of our work, we employ LINE \cite{LINE} as our embedding method to map the user and item relational graphs to their corresponding latent continuous spaces. Note that, one can employ or redesign other embedding methods to create other suitable latent spaces, such as struc2vec \cite{struc2vec}, DeepWalk \cite{DeepWalk}, SDNE \cite{SDNE}. 

Figure~\ref{fig:neighbor}(b1) and (b2) are examples of the latent space after mapping when $l_u = l_v = 2$. Although user $a$ and user $d$ are not in co-occurrence relation, their distance in the latent space may be close because they are the one-hop neighbors of user $b$ and user $c$.

\subsubsection{Step 3: Construction of relational data} 
Based on the latent spaces, user relational data $\mathcal{N}_u$ can be constructed. Specifically, user $u_a$'s relational data $\mathcal{N}_u(a)$ is a user set that contains $K_u$ (a pre-defined hyper-parameter) nearest neighbors in the user latent space based on the particular distance metric in the space.

\par 
Figure~\ref{fig:neighbor}(c1) and (c2) show examples of semantic correlations relation corpus for user $u_a$ and item $v_e$. The neighbor set for user $u_a$ is $\mathcal{N}_u(a) = \left \{  u_b, u_c, u_d \right \}$ when $K_u = 3$, and the neighbor set for item $v_e$ is $\mathcal{N}_v(e) = \left \{  v_b, v_c, v_d \right \}$ when $K_v = 3$. 

\subsection{Recommendation Framework}
\label{sec:framework}
\begin{figure}[h!]
	\centering
	\includegraphics[width=0.94\linewidth]{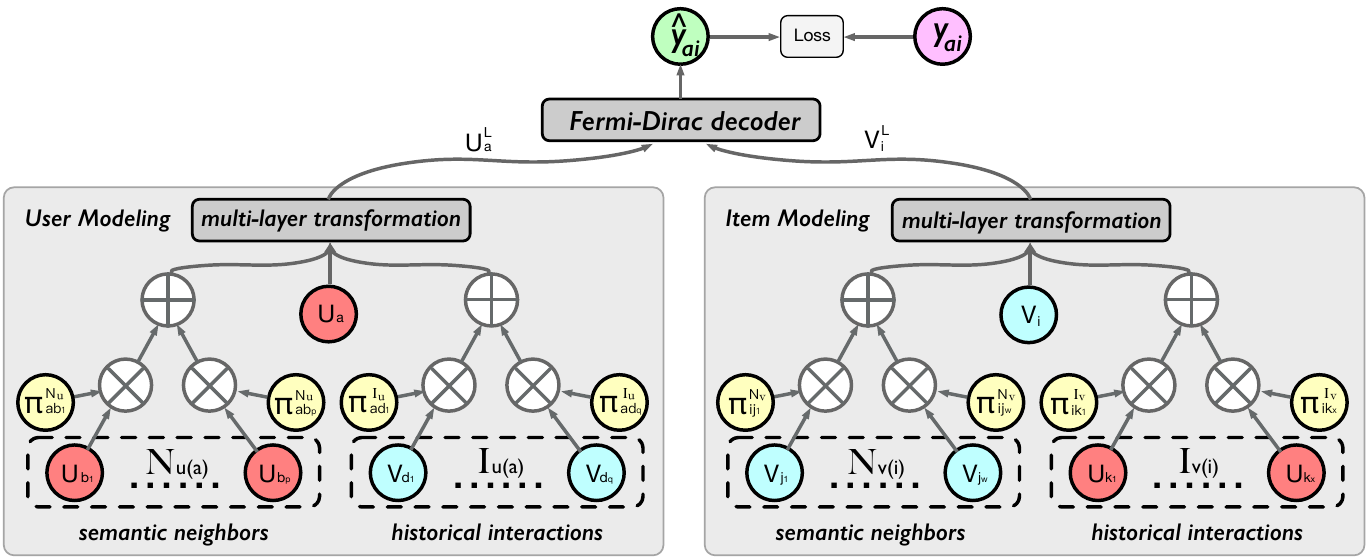}
	\caption{Recommendation framework of HNCR, which is based on hyperbolic representation learning.}
	\label{fig:framework}
\end{figure}

In this subsection, we present the recommendation framework of the HNCR (as illustrated in Figure~\ref{fig:framework}), which is based on hyperbolic representation learning. The framework consists of two parallel neural networks, one for user modeling, and another for item modeling. By taking a user, an item, their semantic neighborhood, and their historical behaviors as inputs, the framework outputs the predicted scores. The details of HNCR are provided as follows.

\subsubsection{Embedding Layer} The embedding layer takes a user and an item as inputs, and encodes them with dense low-dimensional embedding vectors. Specifically, given one hot representations of target user $u_a$ and target item $v_i$, the embedding layer outputs their embeddings $\textbf{u}_{a}$ and $\textbf{v}_{i}$, respectively. We will learn user and item embedding vectors in the hyperbolic space $\mathbb{D}^d_c$.

\subsubsection{Aggregation Layer}
We design an aggregator that aggregates semantic neighborhood and historical behaviors for better modeling user and item embeddings. Since user modeling and item modeling are symmetric, we mainly present user modeling for illustration in the following statement. 

Given user $u_a$, we first aggregate the representations of her semantic neighbors into a single embedding. We can directly utilize M{\"o}bius addition to achieve this goal, as follows:
\begin{equation} 
\begin{split} 
\textbf{u}^{\mathcal{N}}_{a} &= \sum\nolimits_{u_b \in \mathcal{N}_u(a)}^{\oplus_c} {\pi}_{ab}^{\mathcal{N}_u} \otimes_c \textbf{u}_{b},
\label{eq:limit_agg}
\end{split}
\end{equation}
where $\sum\nolimits^{\oplus_c}$ is the accumulation of M{\"o}bius addition, and ${\pi}_{\cdot}^{\cdot}$ is calculated from an attention mechanism which indicates the importance of different neighbors. Specifically, ${\pi}_{ab}^{\mathcal{N}_u}$ is defined as:
\begin{align}
{\pi}_{ab}^{\mathcal{N}_u} = \frac{\exp(- d_c(\textbf{u}_{a}, \textbf{u}_{b})/\tau)}{\sum_{u_{b'} \in \mathcal{N}_u(a)}\exp(- d_c(\textbf{u}_{a}, \textbf{u}_{b'})/\tau) },
\end{align}
where $\tau$ is the temperature parameter which is used for producing a softer distribution over neighbors. Since hyperbolic distance meets the triangle inequality, the attention mechanism can preserve the transitivity among nodes \cite{HNN,HAT}. 

For the second aggregation, it accounts for the user's historical behaviors. Specifically, we aggregate user's interacted items:
\begin{align}
	\textbf{u}^{\mathcal{I}}_{a} = \sum\nolimits_{v_d \in \mathcal{I}_u(a)}^{\oplus_c} {\pi}_{ad}^{\mathcal{I}_u} \otimes_c \textbf{v}_{d},
\label{eq:limit_inter}
\end{align} 
where $\mathcal{I}_u(a)$ is the itemset user $u_a$ shows implicit feedback. ${\pi}_{ad}^{\mathcal{I}_u}$ denotes the attraction of item $d$ to user $a$, which can be defined as:
\begin{align}
{\pi}_{ad}^{\mathcal{I}_u} = \frac{\exp(- d_c(\textbf{u}_{a}, \textbf{v}_{d})/\tau)}{\sum_{v_{d'} \in \mathcal{I}_u(a)}\exp(- d_c(\textbf{u}_{a}, \textbf{v}_{d'})/\tau) }.
\end{align}

The final step in the aggregation layer is to aggregate the target user representation $\textbf{u}_{a}$, her semantic neighborhood representation $\textbf{u}^{\mathcal{N}}_{a}$, and her historical preference representation $\textbf{u}^{\mathcal{I}}_{a}$ into a single vector. We design a multi-layer structure to obtain sufficient representation power, which is formulated as follows: 
\begin{align}
\textbf{u}_{a}^L &= \mathcal{M}^L_u(\mathcal{M}^{L-1}_u(\cdots \mathcal{M}^1_u(\textbf{u}_{a}^0))), \\
\textbf{u}_{a}^0 &= \textbf{u}_{a} \oplus_c\textbf{u}^{\mathcal{N}}_{a} \oplus_c\textbf{u}^{\mathcal{I}}_{a}, \\
\mathcal{M}^l_u(\textbf{u}_{a}^{l-1}) &= \sigma\big ( \textbf{M}^{l}_u \otimes_c \big(\textbf{u}_{a}^{l-1} \oplus_c \textbf{b}^{l}_u\big) \big ), \text{   $l\in \left[1, L\right]$},
\end{align}
where $L$ is the number of hidden layers, $\textbf{M}_u: \mathbb{R}^{d}\rightarrow \mathbb{R}^{d}$ is a linear map, $b_u \in \mathbb{D}^d_c$ is the bias,and $\sigma$ is the nonlinear activation function defined as LeakyReLU \cite{LeakyReLU}.

\subsubsection{Prediction Layer}
After the aggregation layer, we feed user aggregation representation $\textbf{u}_{a}^L$ and target item aggregation representation $\textbf{v}_{i}^L$ into a function $p$ for predicting the probability of $u_a$ engaging $v_i$: $\hat{y}_{ai} = p(\textbf{u}_{a}^L, \textbf{v}_{i}^L)$. Here we implement function $p$ as the Fermi-Dirac decoder \cite{hp1,Poincare}, a generalization of sigmoid function, to compute probability scores between $u_a$ and $v_i$:
\begin{align}
\hat{y}_{ai} = \frac{1}{e^{(d_c(\textbf{u}_{a}^L, \textbf{v}_{i}^L) -r)/t}+1},
\label{eq:predict}
\end{align}
where $r$ and $t$ are hyper-parameters. 

\subsubsection{Framework Optimization}

\paragraph{Objective Function}
To estimate parameters of HNCR's framework, we have the following objective function:
\begin{align} 
\underset{\Theta}{\min}\mathcal{L} = - \!\!\! \sum_{(u, v, \tilde{v})\in\mathcal{D}} \!\! \Big ( y_{uv}\log(\hat{y}_{uv}) + (1-y_{u\tilde{v}})\log(1-\hat{y}_{u\tilde{v}}) \Big ),
\end{align}
where $\Theta $ is the total parameter space, including user embeddings $\left \{ \textbf{u}_i \right \}_{i=1}^{\!|\mathcal{U}|}$, item embeddings $\left \{ \textbf{v}_i \right \}_{i=1}^{\!|\mathcal{V}|}$, and weight parameters of the networks $\big \{\textbf{M}_{u}^l, \textbf{M}_{v}^l, {\forall}l \in \left \{ 1, \cdots, L \right \} \big \}$. $\mathcal{D}$ is the set of training triplets. We donate $\mathcal{I}_u$ is the item set which user $u$ has interacted with, and $\mathcal{D}$ can be defined as:
\begin{align} 
\mathcal{D} = \left \{ (u, v, \tilde{v}) \ | \  u\in\mathcal{U} \land v\in\mathcal{I}_u \land \tilde{v}\in\mathcal{V}\backslash\mathcal{I}_u \right \}.
\label{eq:Dr} 
\end{align}

\paragraph{Gradient Conversion}
Since the Poincar$\acute{\text{e}}$ Ball has a Riemannian manifold structure, we utilize Riemannian stochastic gradient descent (RSGD) to optimize our model \cite{RSGD}. As similar to \cite{Poincare}, the parameter updates are of the following form:
\begin{align} 
\theta_{t+1} = \mathfrak{R}_{\theta_t}(-\eta_t\nabla_R\mathcal{L}(\theta_t)),
\end{align}
where $\mathfrak{R}_{\theta_t}$ denotes a retraction onto $\mathbb{D}$ at $\theta$ and $\eta_t$ denotes the learning rate at time $t$. The Riemannian gradient $\nabla_R$ can be computed by rescaling the Euclidean gradient $\nabla_E$ with the inverse of the Poincar$\acute{\text{e}}$ ball metric tensor as $\nabla_R = \frac{ ( 1-\left \| \theta_t \right \|^2 )^2}{4} \nabla_E$.

\subsection{Discussions}

\subsubsection{Acceleration Strategy}
Since M{\"o}bius addition operation is not commutative nor associative \cite{HNN,HAT}, we have to calculate the accumulation of M{\"o}bius addition by order in Equation (14) (for simplicity, we omit the attention score ${\pi}_\cdot^\cdot$) : 
\begin{equation} 
\begin{split}
\textbf{u}^{0}_{a} = \textbf{u}_{a}&\oplus_c\left(\left((\textbf{u}_{b_1} \oplus_c \textbf{u}_{b_2}) \oplus_c \textbf{u}_{b_3}\right) \oplus_c \cdots \right)\\
&\oplus_c\left(\left((\textbf{v}_{d_1} \oplus_c \textbf{v}_{d_2}) \oplus_c \textbf{v}_{d_3}\right) \oplus_c \cdots \right).
\label{eq:limit_agg2}
\end{split}
\end{equation} 

As is known to all, there exist some active users and popular items that have many interactions in real recommendation scenarios, so the calculation in Equation ~(\ref{eq:limit_inter}) is seriously time-consuming, which will affect the efficiency of our method HNCR. Therefore, it is necessary to devise a new way to calculate the aggregation. 

\par
\begin{figure}[h!]
	\centering
	\includegraphics[width=0.98\linewidth]{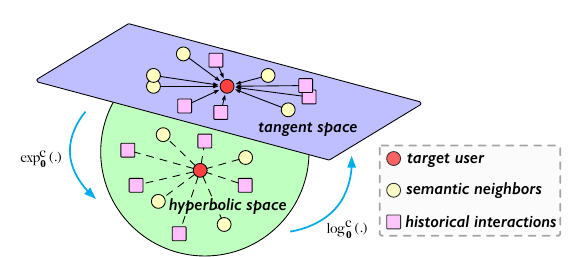}
	\caption{Illustration of the acceleration strategy.}
	\label{fig:map}
\end{figure}

  
Following the approaches in \cite{HAT}, we resort to M{\"o}bius logarithmic map and exponential map, as illustrated in Figure~\ref{fig:map}. Specifically, we first utilize the logarithmic map to project user and item representations into a tangent space, then perform the accumulation operation to aggregate the representations in the tangent space, and finally project aggregated representations back to the hyperbolic space with the exponential map. Take user $u_a$ as an example, the process is formulated as:
\begin{equation} 
\begin{split}
\textbf{u}^{0}_{a} = \exp_{\textbf{0}}^c \Big ( &\log_{\textbf{0}}^c(\textbf{u}_{a}) + \sum\nolimits_{u_b \in \mathcal{N}_u(a)} {\pi}_{ab}^{\mathcal{N}_u} \cdot \log_{\textbf{0}}^c(\textbf{u}_b)  \\
& + \sum\nolimits_{v_d \in \mathcal{I}_u(a)} {\pi}_{ad}^{\mathcal{I}_u} \cdot \log_{\textbf{0}}^c(\textbf{v}_d)\Big ).
\label{eq:eff_agg}
\end{split}
\end{equation}

Different from Equation (14), we can calculate the results in a parallel way in Equation ~(\ref{eq:eff_agg}) because the accumulation operation in the tangent space is commutative and associative, which enables our model more efficient. Therefore, we replace Equation (14) with Equation ~(\ref{eq:eff_agg}) for neighbor aggregation.

\subsubsection{Time Complexity Analysis} 

The time cost of our proposed method HNCR is from two parts: neighbor construction and recommendation framework. (i) For the neighbor construction, the time complexity is $O(N\cdot C_u + M\cdot C_v)$, where $C_u$ and $C_v$ denote the average number of co-occurrence neighbors for users and items, respectively. For the relational graph mapping, the time complexity depends on the selected node embedding method. For the process of obtaining semantic neighbors, the time complexity for a user is $O(M\cdot l_{u})$. Similarly, for an item, the time complexity is $O(N\cdot l_{v})$. In practice, we can utilize some acceleration computation methods proposed by previous works \cite{KDT,BT1,BT2} to speed up the process of obtaining semantic neighbors. Note that the relational data $\mathcal{N}_u$ and $\mathcal{N}_v$ can be computed offline in advance, so we can prepare $\mathcal{N}_u$ and $\mathcal{N}_v$ before generating recommendations. (ii) As we can see, the neighbor aggregation is the main operation in recommendation framework. For a user, the time consumption of the attention mechanism is $O(K_u \cdot d + H_u \cdot d)$, where $K_u$ is the number of semantic neighbors for each user, $H_u$ denotes the average number of user's interaction, and $d$ denotes the embedding size. The time cost of the multi-layer structure is $O(L\cdot d^2)$, where $L$ is the total layers in the aggregation layer. Similarly, the time consumption for an item in the aggregation layer is $O(K_v \cdot d + H_v \cdot d + L\cdot d^2)$, where $K_u$ is the number of semantic neighbors for each item and $H_v$ denotes the average number of items' interactions. In general, the time consumption of the whole training epoch is $O( Y \cdot ((K_u + H_u + K_v + H_v) \cdot d + L\cdot d^2) )$, where $Y$ denotes the number of user-item interactions. 

\section{Experiments}
\label{sec:experiment}

\subsection{Experiment Setup}

\subsubsection{Datasets} 
We experimented with four datasets: Ciao\footnote{Ciao: http: //www.cse.msu.edu/\textasciitilde tangjili/index.html}, Yelp\footnote{Yelp: http://www.yelp.com/}, Epinion\footnote{Epinion: http://alchemy.cs.washington.edu/data/epinions/}, and Douban\footnote{Douban: http://book.douban.com}. Each dataset contains users’ ratings of the items. In the data preprocessing step, we transform the ratings into implicit feedback (denoted by “1”) indicating that the user has rated the item positively. Then, for each user, we sample the same amount of negative samples (denoted by “0”) as their positive samples from unwatched items. The statistics of the datasets are summarized in Table~\ref{exp:table1}.

\begin{table}[h!]
\begin{spacing}{0.8}
\centering  
\caption{Statistical details of the four datasets.}
\begin{tabular}{{c|c c c c}}
	\hline
	dataset  & \# users & \# items & \# interactions & density  \\ \hline\hline
	Ciao     & 7,267  & 11,211 & 147,995 & 0.181\%  \\ \hline
	Yelp     & 10,580 & 13,870 & 171,102 & 0.116\%  \\ \hline
	Epinion  & 20,608 & 23,585 & 454,022 & 0.093\%  \\ \hline
	Douban   & 12,748 & 22,347 & 785,272 & 0.275\%  \\ \hline
\end{tabular}
\label{exp:table1}
\end{spacing}
\end{table}

\begin{table*}[!]
\begin{spacing}{0.95}
	\centering 
	\caption{The results of \textit{AUC} and \textit{Accuracy} in CTR prediction on four datasets. ** denotes the best values among all methods, and * denotes the best values among all competitors.}
	\begin{tabular}{ c | c c | c c | c c | c c }
		\hline
		\multirow{2}{*}{Method}&\multicolumn{2}{c|}{Ciao}&\multicolumn{2}{c|}{Yelp}&\multicolumn{2}{c|}{Epinion}&\multicolumn{2}{c}{Douban}\cr\cline{2-3}\cline{4-5}\cline{6-7}\cline{8-9}
		&AUC&ACC&AUC&ACC&AUC&ACC&AUC&ACC \cr
		\hline\hline
		SVD     & 0.7240 & 0.6574 & 0.8170 & 0.7518 & 0.8026 & 0.7279 & 0.8257 & 0.7514 \\
		NFM     & 0.7333 & 0.6638 & 0.8230 & 0.7588 & 0.8094 & 0.7330 & 0.8373 & 0.7601 \\
		CMN     & 0.7360 & 0.6613 & 0.8255 & 0.7595 & 0.8133 & 0.7393 & 0.8382 & 0.7597 \\
		MMCF    & 0.7567 & 0.6838 & 0.8284 & 0.7635 & 0.8269 & 0.7486 & 0.8498 & \textbf{ 0.7728*} \\
		NGCF    & 0.7646 & 0.6948 & 0.8279 & 0.7628 & 0.8235 & 0.7450 & 0.8512 & 0.7721 \\
		LR-GCCF & \textbf{ 0.7683*} & \textbf{ 0.6980*} & \textbf{ 0.8354*} & \textbf{ 0.7665*} & \textbf{ 0.8305*}& \textbf{ 0.7541*} & \textbf{ 0.8521*} & 0.7717 \\
		Poincar$\acute{\text{e}}$Emb & 0.7595 & 0.6880 & 0.8230 & 0.7608 & 0.8029 & 0.7342 & 0.8377 & 0.7669 \\\hline 
		HNCR    & \textbf{  0.8002**} & \textbf{  0.7104**} & \textbf{  0.8598**} & \textbf{  0.7915**} & \textbf{  0.8527**} & \textbf{  0.7703**} & \textbf{  0.8792**} & \textbf{  0.8018**}\\
		ENCR    & 0.7763 & 0.6985 & 0.8295 & 0.7650 & 0.8301 & 0.7518 & 0.8441 & 0.7690 \\
		\hline
	\end{tabular}
	\label{exp:table2}
\end{spacing}
\end{table*}

\subsubsection{Comparison Methods}
To verify the performance of our proposed method HNCR, we compared it with the following state-of-art recommendation methods. The characteristics of the comparison methods are listed as follows:
\begin{itemize}[leftmargin= 13 pt]
\item\textbf{SVD} is a famous baseline which is a hybrid model combining the latent factor model and the neighborhood model \cite{SVD}.
\item\textbf{NFM} is a feature-based factorization model, which improves FM \cite{FM} by using the MLP component to capture high-order feature interaction \cite{NFM}. Here we concatenate user ID embedding and item ID embedding as input for NFM.
\item\textbf{CMN} is a memory-based model, which designs the memory slots of similar users to learn user embeddings \cite{CMN}. Note that it only focuses on the user's neighbors without accounting for the information about similar items.
\item\textbf{MMCF} is a memory-based model, which models user-user and item-item co-occurrence contexts by memory networks \cite{MMCF}. Different from our methods, it only focuses on co-occurrence relations and ignores high-order semantic transitive relations.
\item\textbf{NGCF} is a graph-based recommender system, which utilizes multiple propagation layers to learn user and item representations by propagating embeddings on the bipartite graph \cite{NGCF}. 
\item\textbf{LR-GCCF} is a graph-based recommender system, which designs a linear propagation layer to leverage the user-item graph structure for user and item embedding modeling \cite{LRGCCF}.
\item\textbf{Poincar$\acute{\text{e}}$Emb} is a hyperbolic embedding method \cite{Poincare}. Here Poincar$\acute{\text{e}}$Emb considers matrix completion for recommendation from the point of view of link prediction on graphs.
\item\textbf{ENCR} is the Euclidean counterpart of HNCR, which replaces M{\"o}bius addition, M{\"o}bius matrix-vector multiplication, Gyrovector space distance with Euclidean addition, Euclidean matrix multiplication, Euclidean distance, and remove M{\"o}bius logarithmic map and exponential map.    
\item\textbf{HNCR} is our complete model.
\end{itemize}

It is worth noting that, MMCF, NGCF, LR-GCCF are recently proposed state-of-the-art models.

\subsubsection{Parameter Settings}
\par 
We implemented our method using the python library of Pytorch. For each dataset, we randomly split it into training, validation, and test sets following 6 : 2 : 2. The learning rate $\eta$ is tuned among $[10^{-4}, 5\times10^{-4}, 10^{-3}, 5\times10^{-3}, 10^{-2}]$; the embedding size $d$ is searched in $[8, 16, 32, 64, 128]$; the semantic neighbor size factor $K_u,K_v$ is chosen from $[5, 10, 15, 20, 25]$; and the layer size $L$ is selected from $[1, 2, 3, 4]$. In addition, we set batch size $b=1024$, curvature $c=1$, temperature $\tau=0.1$, and Fermi-Dirac decoder parameters $r = 2$, $t = 1$. The best settings for the hyper-parameters in all baselines are reached by either empirical study or following their original papers.
\par

\subsubsection{Evaluation Protocols} 
We evaluate our method HNCR in two experiment scenarios: (i) in click-through rate (CTR) prediction, we adopt two metrics \textit{AUC} (area under the curve) and \textit{Accuracy}, which are widely utilized in binary classification problems; and (ii) in top-$K$ recommendation, we use the model obtained in CTR prediction to generate top-$K$ items. Since it is time-consuming to rank all items for each user in the evaluation procedure, to reduce the computational cost, following the strategy in \cite{NCF,Diffnet}, for each user, we randomly sample 1000 unrated items at each time and combine them with the positive items in the ranking process. We use the metrics \textit{Precision@K} and \textit{Recall@K} to evaluate the recommended sets. We repeated each experiment 5 times and reported the average performance.

\subsection{Empirical Study}
\begin{figure}[h!]
	\centering
	\includegraphics[width=0.9\linewidth]{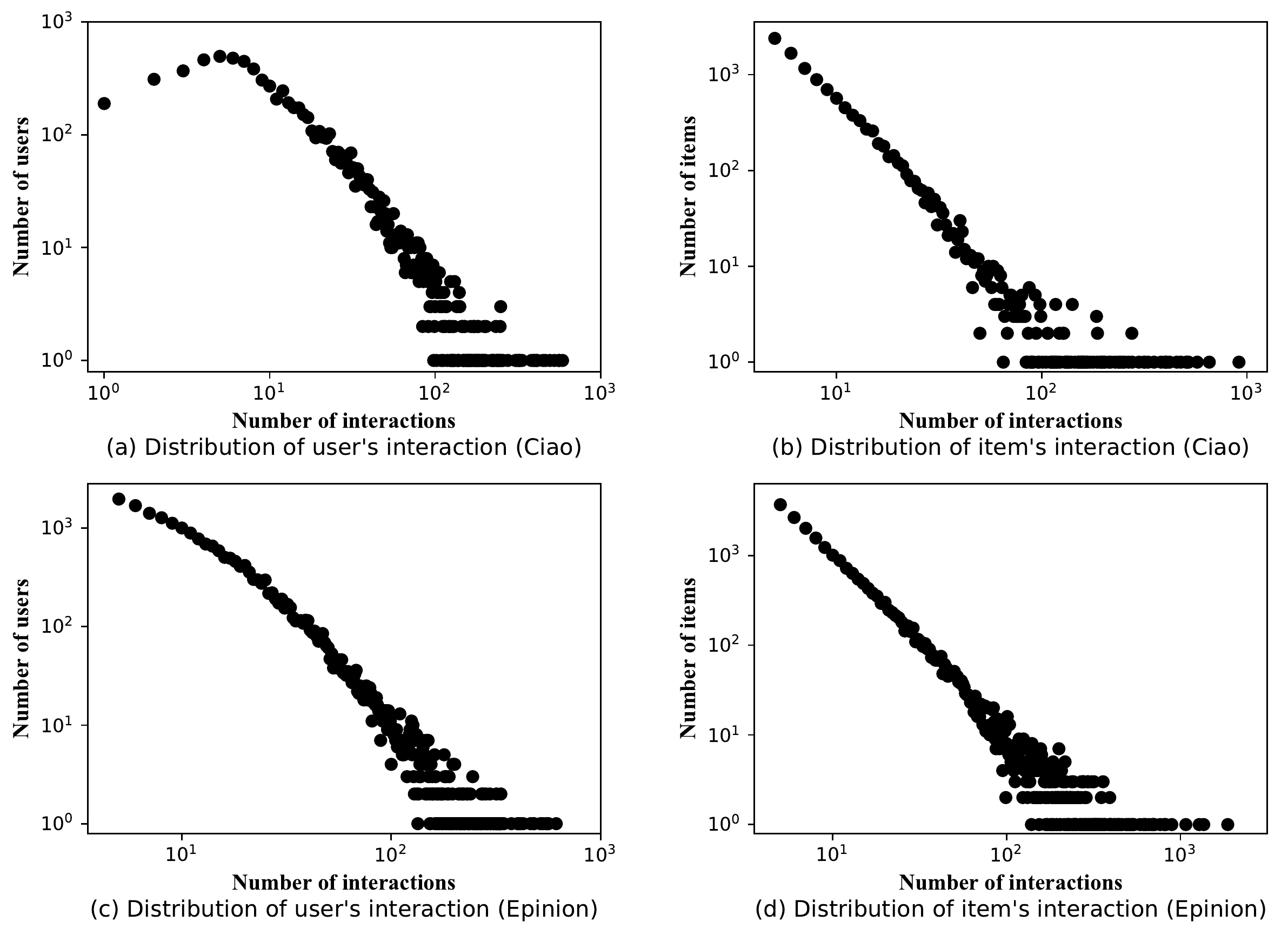}
	\caption{Distributions of the user-item interaction on Ciao (top row) and Epinion (bottom row). The X-axis presents the number of interactions associated with a user or item, and the Y-axis shows the number of such users or items.}
	\label{fig:powlaw} 
\end{figure}

Researches show that data with a power-law structure can be naturally modeled in the hyperbolic space \cite{HHNE,Poincare,hp1}. Therefore, we conduct an empirical study to check whether the power-law distribution also exists in the user-item interaction relation. We present the distribution of the number of interactions for users and items in Figure~\ref{fig:powlaw}. Due to the space limitation, we only show the results of Ciao and Epinion datasets. We observed that these distributions show the power-law distribution: a majority of users/items have very few interactions, and a few users/items have a huge number of interactions. The above findings empirically demonstrate user-item interaction relations exhibit power-law structure, thus we believe that using hyperbolic geometry might be suitable for the CF task.

\begin{figure*}[h!]
	\centering
	\includegraphics[width=0.98\linewidth]{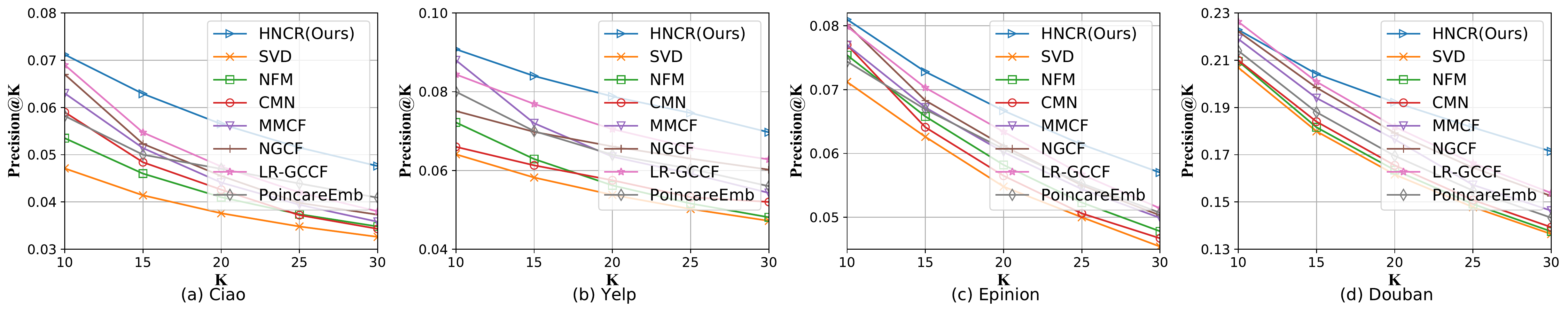}
	\caption{The results of \textit{Precision@K} in top-$K$ recommendation on four datasets.}
	\label{fig:pre}
\end{figure*}

\begin{figure*}[h!]
	\centering
	\includegraphics[width=0.98\linewidth]{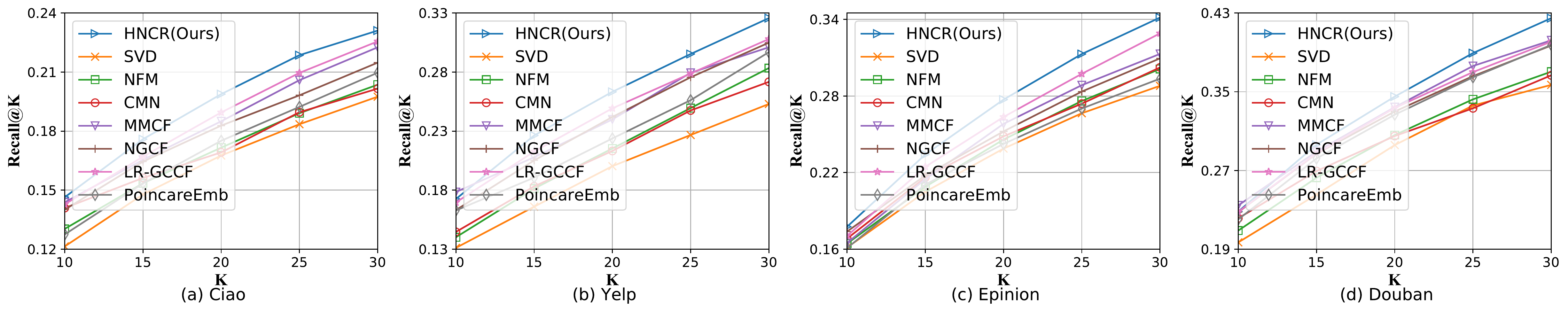}
	\caption{The results of \textit{Recall@K} in top-$K$ recommendation on four datasets.}
	\label{fig:rec}
\end{figure*}

\subsection{Performance Comparison}
Table~\ref{exp:table2} and Figures~\ref{fig:pre}, ~\ref{fig:rec} show the performance of all compared methods in CTR prediction and top-$K$ recommendation (ENCR are not plotted in Figure 2 for clarity), respectively. From the results, we have the following main observations:
\par
(i) SVD achieves poor performance on four datasets, which indicates the shallow representation is insufficient to capture complex user-item interaction. NFM consistently outperforms SVD, which suggests the significance of non-linear feature interactions between user and item embeddings in recommender systems. However, both SVD and NFM ignore user-user and item-item relations.

(ii) CMN and MMCF generally achieve better performance than NFM in most cases. This may because both of them 
consider relations among users (or items). Besides, MMCF consistently outperforms CMN. It makes sense since CMN only account for user neighbor information, while MMCF considers co-occurrence information for both users and items.

(iii) Both SVD and Poincar$\acute{\text{e}}$Emb are shallow representation models, while Poincar$\acute{\text{e}}$Emb achieves better performance; meanwhile, HNCR consistently outperforms Euclidean variant ENCR. These results indicate that using hyperbolic space for learning user-item embeddings can enhance the recommendation performance. 

(iv) Intuitively, HNCR has made great improvements over state-of-the-art baselines in both recommendation scenarios. For CTR prediction task, our method HNCR yields the best performance on four datasets. For example, HNCR improves over the strongest baselines \textit{w.r.t.} \textit{AUC} by 4.15\%, 2.92\%, 2.67\%,and 3.18\% in Ciao, Yelp, Epinion and Douban datasets, respectively. In top-$K$ recommendation, HNCR achieves 4.91\%, 5.61\%, 5.15\%, and 3.47\% performance improvement against the strongest baseline \textit{w.r.t.} \textit{Recall@20} in Ciao, Yelp, Epinion and Douban datasets, respectively.

\subsection{Handling Data Sparsity Issue}
\begin{figure}[h!]
	\centering
	\includegraphics[width=0.98\linewidth]{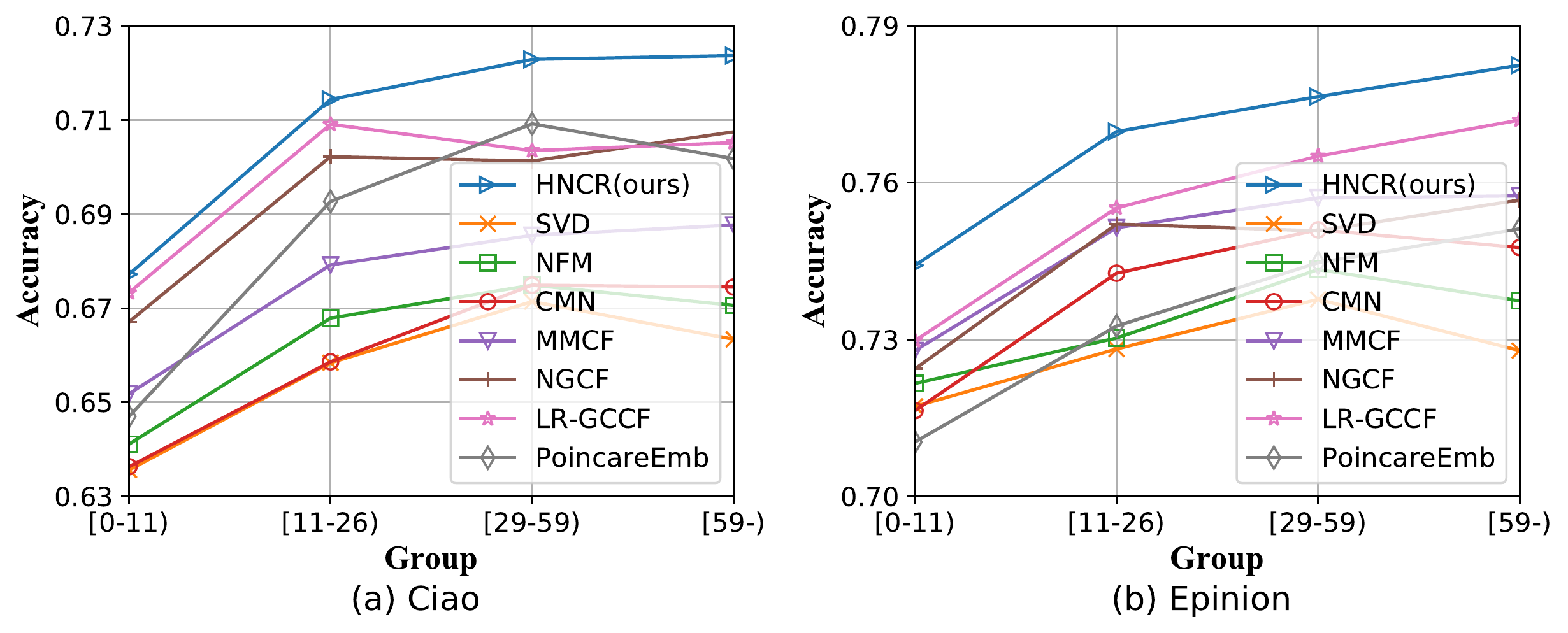}  
	\caption{Performance comparison over the sparsity distribution of user groups on Ciao and Epinion datasets.}
	\label{fig:sparse}
\end{figure}

The data sparsity problem is a great challenge for most recommender systems. To investigate the effect of data sparsity, we bin the test users into four groups with different sparsity levels based on the number of observed ratings in the training data, meanwhile, keep each group including a similar number of interactions. For example, [11,26) in the Ciao dataset means for each user in this group has at least 11 interaction records and less than 26 interaction records. Due to the space limitation, we show the \textit{Accuracy} results on different user groups with different models of Ciao and Epinion datasets in Figure~\ref{fig:sparse}. From the results, we observe that HNCR consistently outperforms the other methods including the state-of-the-art methods like MMCF and LR-GCCF, which verifies our method can maintain a decent performance in different sparse scenarios. 

\subsection{Ablation Study}
\subsubsection{Effect of Weighted Strategy}
To explore the effect of our weighted strategy in relational graph construction, we conducted experiments with the two variants of HNCR and ENCR: (1) HNCR-N and ENCR-N (using the number of common neighbors in the bipartite graph as the weight), and (2) HNCR-0 and ENCR-0 (without using any weighted strategy). Table~\ref{exp:weight} shows the \textit{AUC} results on four datasets. From the results, we find using the other two methods leads to a slight decrease in performance. Although the performance drop is not quite significant, the experiment shows that our weighted strategy is beneficial.

\begin{table}[h!]
\begin{spacing}{0.9}
\centering
\caption{Effect of the weighted strategy on four datasets.}
\begin{tabular}{ c | c | c | c | c }  %
\hline 
Dataset   & Ciao   & Yelp   & Epinion & Douban  \\  \hline
HNCR      & 0.8002 & 0.8598 & 0.8527  & 0.8792  \\  
HNCR-N    & 0.7986 & 0.8566 & 0.8515  & 0.8770  \\  
HNCR-0    & 0.7972 & 0.8570 & 0.8499  & 0.8772  \\  \hline 
ENCR      & 0.7763 & 0.8295 & 0.8301  & 0.8441  \\  
ENCR-N    & 0.7727 & 0.8282 & 0.8295  & 0.8418  \\  
ENCR-0    & 0.7722 & 0.8274 & 0.8271  & 0.8407  \\  \hline 
\end{tabular}
\label{exp:weight}
\end{spacing}
\end{table}

\subsubsection{Semantic Neighborhood vs Co-occurrence Neighborhood}
To compare the effectiveness of semantic neighborhood and co-occurrence neighborhood, we design two variants HNCR-C and ENCR-C by replacing semantic neighborhood with co-occurrence neighborhood in the aggregation layer. We also design a variant MMCF-S for MMCF, in which it utilizes our constructed semantic neighborhood instead of co-occurrence neighborhood. Table~\ref{exp:vs} shows the \textit{AUC} results on four datasets. From the results, we find HNCR-C and ENCR-C perform worse than HNCR and ENCR, respectively, and MMCF-S achieve better performance than MMCF, which verifies constructed semantic neighbors can provide more useful information than co-occurrence neighborhood.

\begin{table}[h!]
\begin{spacing}{0.9}
\centering
\caption{Effect of semantic neighborhood and co-occurrence neighborhood on four datasets.}
\begin{tabular}{ c | c | c | c | c }  %
\hline 
Dataset   & Ciao   & Yelp   & Epinion & Douban  \\  \hline
HNCR      & 0.8002 & 0.8598 & 0.8527  & 0.8792  \\  
HNCR-C    & 0.7933 & 0.8510 & 0.8468  & 0.8704  \\  \hline 
ENCR      & 0.7763 & 0.8295 & 0.8301  & 0.8441 \\  
ENCR-C    & 0.7698 & 0.8211 & 0.8237  & 0.8366 \\  \hline 
MMCF      & 0.7567 & 0.8284 & 0.8269  & 0.8498  \\  
MMCF-S    & 0.7651 & 0.8346 & 0.8315  & 0.8557  \\  \hline 
\end{tabular}
\label{exp:vs}
\end{spacing}
\end{table}

\subsubsection{Effect of Aggregator}
The key part of the recommendation framework is that we devise an aggregator to refine user and item hyperbolic representations. In this subsection, we evaluate the aggregator by analyzing the contributions from different components. 

To this end, we conducted experiments with the three variants of HNCR and ENCR: (1) HNCR-S and ENCR-S (without using semantic neighbor information); (2) HNCR-H and ENCR-H (without using historical behavior information); and (3) HNCR-A and ENCR-A (without using attention mechanism). Table~\ref{exp:agg} shows the \textit{AUC} results of different variants on four datasets. From the results, we find that removing any components will decrease recommendation performance of our models. For example, HNCR-S and HNCR-H perform worse than the complete model HNCR, which shows that both semantic neighbors and historical behaviors benefit the recommendation; HNCR also achieves better scores than HNCR-A, which validates considering the importance of different neighbors in aggregation operation is helpful for improving performance.

\begin{table}[h!]
\begin{spacing}{0.9}
\centering
\caption{Effect of aggregator's components on four datasets.}
\begin{tabular}{ c | c | c | c | c }  %
\hline 
Dataset   & Ciao   & Yelp   & Epinion & Douban  \\  \hline
HNCR      & 0.8002 & 0.8598 & 0.8527  & 0.8792  \\  
HNCR-S    & 0.7918 & 0.8488 & 0.8432  & 0.8685  \\  
HNCR-H    & 0.7880 & 0.8462 & 0.8441  & 0.8673  \\  
HNCR-A    & 0.7943 & 0.8521 & 0.8494  & 0.8746  \\  \hline 
ENCR      & 0.7763 & 0.8295 & 0.8301  & 0.8441  \\  
ENCR-S    & 0.7644 & 0.8187 & 0.8210  & 0.8322  \\  
ENCR-H    & 0.7621 & 0.8120 & 0.8184  & 0.8301  \\  
ENCR-A    & 0.7745 & 0.8286 & 0.8297  & 0.8404  \\  \hline 
\end{tabular}
\label{exp:agg}
\end{spacing}
\end{table}

\subsection{Parameter Sensitivity}
\begin{figure}[h!]
	\centering
	\includegraphics[width=0.98\linewidth]{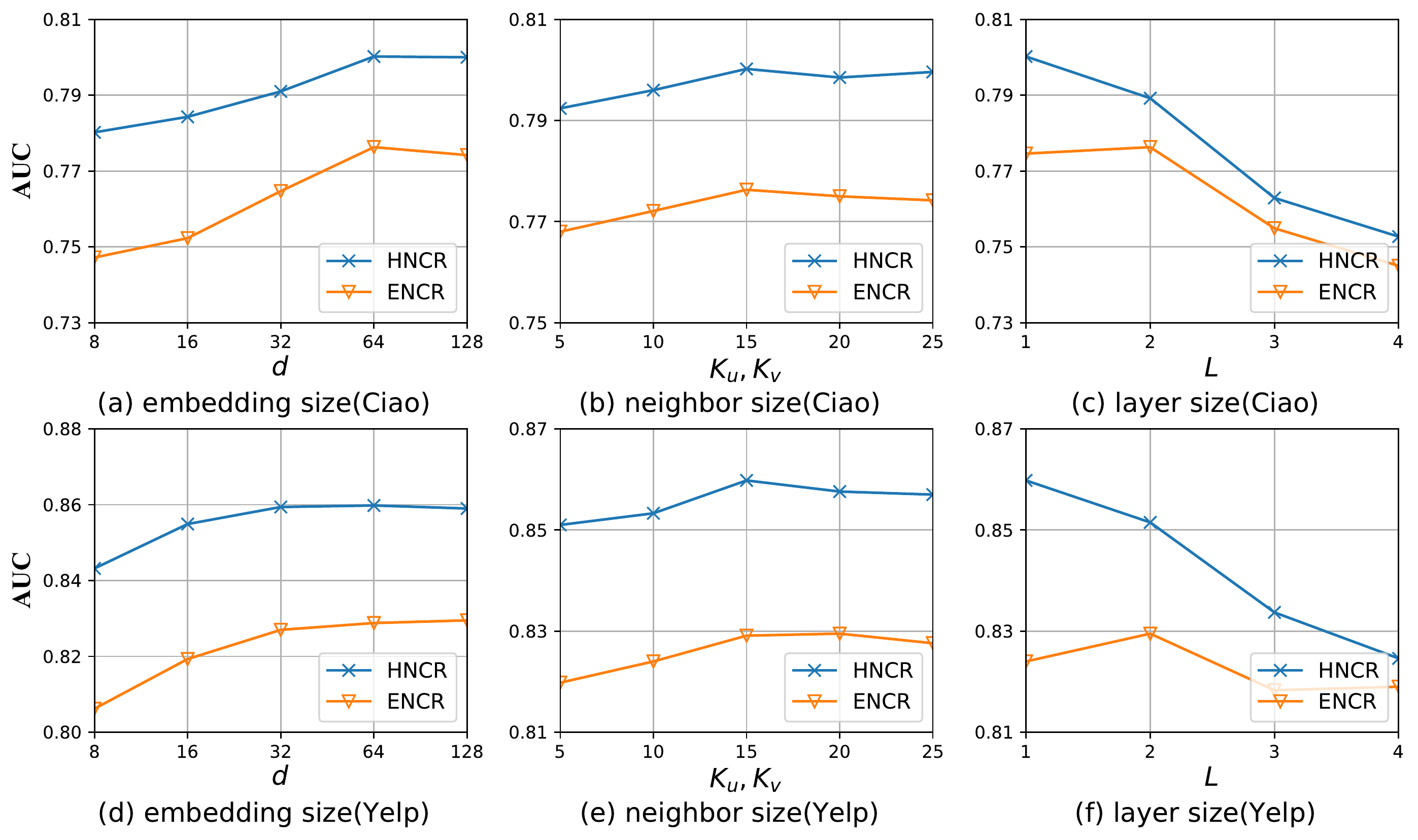}  
	\caption{Parameter sensitivity on Ciao and Yelp.}
	\label{fig:parameters}
\end{figure}

We explore the impact of three hyper-parameters: embedding size $d$, semantic neighbor size $K_u,K_v$, and layer size $L$. The results on Ciao and Yelp are plotted in Figure~\ref{fig:parameters}. We have the following observations: (i) A proper embedding size $d$ is needed. If it is too small, the model lacks expressiveness, while a too large $d$ increases the complexity of the recommendation framework and may overfit the datasets. In addition, we observe that HNCR always significantly outperforms ENCR regardless of the embedding size, especially in a low-dimensional space, which shows that HNCR can effectively learn high-quality representations for CF tasks. (ii) For neighbor size $K_u,K_v$, we find that the \textit{AUC} results increase first and then start to decrease. It may probably because a larger semantic neighbor size will be prone to be misled by noises. (iii) For layer size $L$, we observe HNCR and ENCR achieve the best performance when $L=1$ and $L=2$, respectively, which shows that stacking more layers are not helpful in our models. The results also indicate that utilizing hyperbolic geometry can achieve good performance without designing complex multi-layer structures.

\subsection{Case Study}
In the empirical study, we have shown that the user-item interaction relation exhibits the power-law structure. Power-law distributions often suggest the underlying hierarchical structure \cite{HOCN,Poincare}. In this case study, we evaluate whether learned embeddings in our models can reflect such structure in the user-item bipartite graph. 

\begin{figure}[h!]
	\centering
	\includegraphics[width=0.98\linewidth]{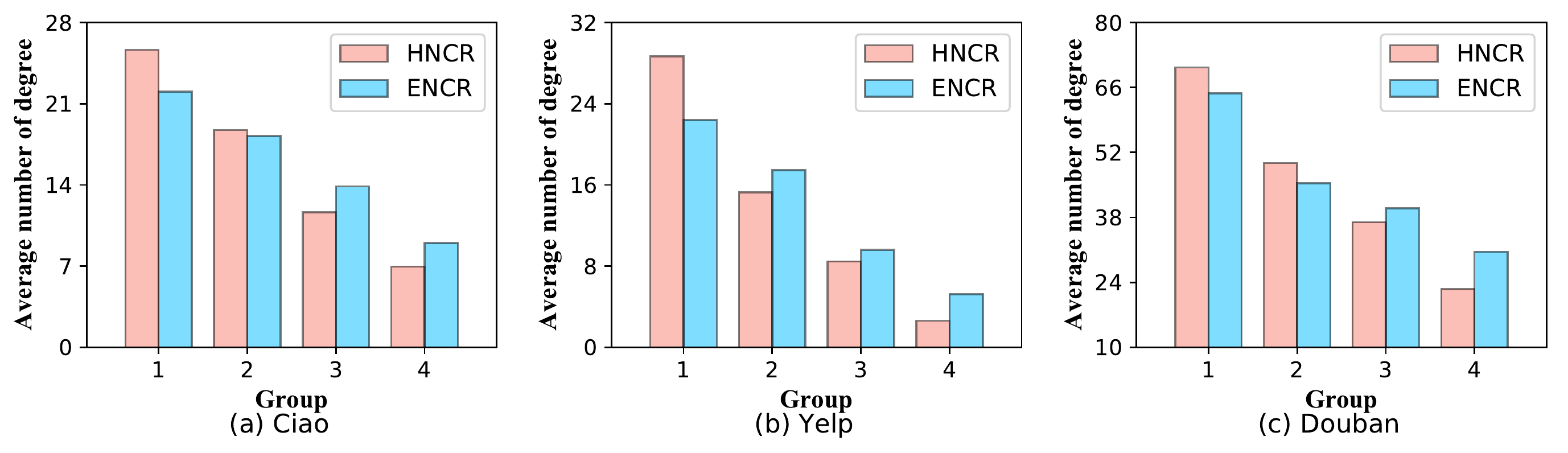}  
	\caption{Analysis of hierarchical structure on Ciao, Yelp, and Douban datasets.}
	\label{fig:hierarchical1}
\end{figure}

In general, the distances between embeddings and the origin can reflect the latent hierarchy of graphs \cite{Poincare,HHNE}. We utilize Gyrovector space distance and Euclidean distance to calculate the distance to the origin for HNCR and ENCR, respectively. We bin the nodes in the user-item bipartite graph into four groups according to their distances to the origin (from the near to the distant), meanwhile, keep each group including a similar number of nodes. For example, nodes in group 1 have the nearest distances to the origin while nodes in group 4 have the furthest distances to the origin. To evaluate the nodes' activity in the bipartite graph, we compute the average number of nodes' interaction behaviors in each group. Figure~\ref{fig:hierarchical1} shows the results on Ciao, Yelp, and Douban datasets. From the results, we can see that the average number of interaction behaviors decreases from group 1 to group 4. This result indicates that hierarchy of interaction behaviors can be modeled by our methods HNCR and ENCR. Compared with ENCR, we find that HNCR more clearly reflects the hierarchical structure, which indicates that hyperbolic space is more suitable than Euclidean space to embed data with the power-law distribution.

\begin{figure}[h!]
	\centering
	\includegraphics[width=0.98\linewidth]{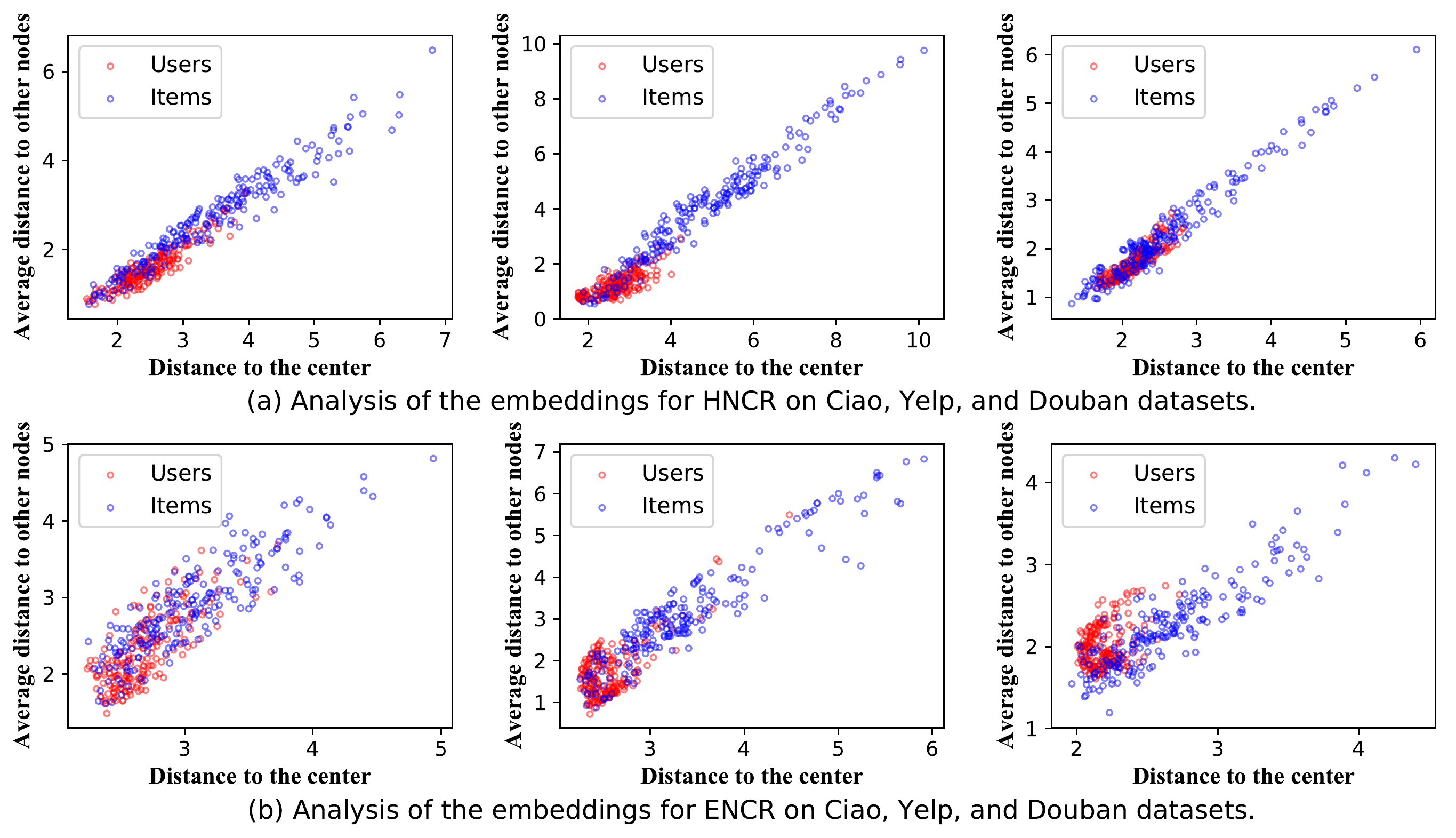}  
	\caption{Analysis of the embeddings for our methods on Ciao, Yelp, and Douban datasets.}
	\label{fig:hierarchical2}
\end{figure}

To further show the hierarchical structure learned in the embeddings, we randomly select 200 users and 200 items from Ciao, Yelp, and Douban datasets, and plot them in Figure~\ref{fig:hierarchical2} (from left to right) where the x-axis and y-axis represent the distance from the center and the average distance from all other nodes in the dataset, respectively. The results show that active nodes (near the center) generally have small average distances and vice versa. Moreover, we find that users are closer to the center than items, which is consistent with previous studies \cite{HyperML}. Compared with ENCR, we find the distribution of nodes in HNCR is more regular, which indicates utilizing hyperbolic geometry to learn user and item representations can better organize the underlying hierarchical structure.

\section{Related Work}
\label{sec:related}
In this section, we provide a brief overview of two areas that are highly relevant to our work.

\subsection{Collaborative Filtering}
Collaborative filtering can generally be grouped into three categories: neighborhood-based model, latent factor model, and hybrid model \cite{RecBook,SVD}. Neighborhood-based methods are centered on identifying neighborhoods of similar users or items based on the user-item interaction history \cite{item_item,item_item2}. For example, ItemKNN utilizes collaborative item-item similarities (i.g. cosine similarity) to generate recommendations \cite{item_item}. The latent factor model, such as Matrix Factorization\cite{MF}, projects users and items into low-dimensional vector spaces and then uses the inner product to model the interactions. With the development of deep learning, some latent factor models utilize deep neural networks as representation learning tools to capture complex user-item interactions \cite{NCF,DMF,CDAE}. To further capture high-order interaction information between users and items, researchers propose using GCNs and their variants in CF tasks \cite{NGCF,LightGCN,GCCF,SpectralCF,KNI,LRGCCF}. As for the hybrid model, it merges the latent factor model and the neighborhood-based model. SVD is a well-known hybrid model, which leverages users’ explicit feedbacks and implicit feedbacks to predict user preferences \cite{SVD}. Recently, a line of work leverages co-occurrence relations to define the neighbors for users and items and integrates deep components into the hybrid model \cite{MMCF,CMN}. Since the co-occurrence relation is coarse-grained and lacks high-order semantics, these methods are insufficient to generate better recommendations. Different from the above-mentioned work, our method HNCR first devises a neighbor construction strategy to find semantic neighbors for users and items and then carries a deep framework based on hyperbolic representation learning to integrate constructed neighbor sets and historical behaviors into recommendation.

\subsection{Hyperbolic Representation Learning}
In recent years, representation learning in hyperbolic spaces has attracted an increasing amount of attention. Specifically, \cite{Poincare} embedded hierarchical data into the Poincar$\acute{\text{e}}$ ball, showing that hyperbolic embeddings can outperform Euclidean embeddings in terms of both representation capacity and generalization ability. \cite{hyperrelated1} focused on learning embeddings in the Lorentz model and showed that the Lorentz model of hyperbolic geometry leads to substantially improved embeddings. \cite{hyperrelated2} extended Poincar$\acute{\text{e}}$ embeddings to directed acyclic graphs by utilizing hyperbolic entailment cones. \cite{hyperrelated3} analyzed representation trade-offs for hyperbolic embeddings and developed and proposed a novel combinatorial algorithm for embedding learning in hyperbolic space. Besides, researchers began to combining hyperbolic embedding with deep learning. \cite{HNN} introduced hyperbolic neural networks which defined core neural network operations in hyperbolic space, such as M{\"u}bius addition, M{\"u}bius scalar multiplication, exponential and logarithmic maps. After that, hyperbolic analogues of other algorithms have been proposed, such as Poincar$\acute{\text{e}}$ Glove \cite{PoincareGlove}, hyperbolic graph attention network \cite{HAT}, and hyperbolic attention networks \cite{hp2}.  

Some recent works using hyperbolic representation learning for CF tasks. For instance, HyperBPR learns user and item hyperbolic representations and leverages Bayesian Personalized Ranking (BPR) for recommendation \cite{HyperBPR}. HyperML is a metric learning method, which makes use of M{\"o}bius gyrovector space operations to design the distance of user-item pairs \cite{HyperML}. However, these models are different from our model as they treat every user-item pair as an isolated data instance, without considering semantic relations among users (or items). Also, they are the shallow models, which may lack expressiveness to model features for users and items. In this paper, we explicitly consider semantic correlations among users/items and develop a deep framework based on hyperbolic representation learning for recommendation. To the best of our knowledge, this work is the first hyperbolic neural approach that explicitly models user-user and item-item semantic relations for CF tasks.

\section{Conclusion and Future Work}
\label{sec:conclusion}
In this work, we make use of hyperbolic geometry and deep learning techniques for recommendation. We develop a novel method called HNCR, which includes (i) a neighbor construction method that utilizes the user-item interaction information to construct semantic neighbor sets for users and items; and (ii) a deep framework that uses hyperbolic geometry to integrate constructed neighbor sets and interaction history into recommendation. Extensive experimental results on four datasets demonstrate HNCR outperforms its Euclidean counterpart and state-of-the-art models.

For future work, we will (i) integrate side information into HNCR such as knowledge graphs and social networks to further enhance the performance; and (ii) try to generate recommendation explanations for comprehending the user behaviors and item attributes.

\newpage

\bibliographystyle{ACM-Reference-Format}
\bibliography{HNCR}


\begin{thebibliography}{50}


\ifx \showCODEN    \undefined \def \showCODEN     #1{\unskip}     \fi
\ifx \showDOI      \undefined \def \showDOI       #1{#1}\fi
\ifx \showISBNx    \undefined \def \showISBNx     #1{\unskip}     \fi
\ifx \showISBNxiii \undefined \def \showISBNxiii  #1{\unskip}     \fi
\ifx \showISSN     \undefined \def \showISSN      #1{\unskip}     \fi
\ifx \showLCCN     \undefined \def \showLCCN      #1{\unskip}     \fi
\ifx \shownote     \undefined \def \shownote      #1{#1}          \fi
\ifx \showarticletitle \undefined \def \showarticletitle #1{#1}   \fi
\ifx \showURL      \undefined \def \showURL       {\relax}        \fi
\providecommand\bibfield[2]{#2}
\providecommand\bibinfo[2]{#2}
\providecommand\natexlab[1]{#1}
\providecommand\showeprint[2][]{arXiv:#2}

\bibitem[\protect\citeauthoryear{Bentley}{Bentley}{1975}]%
        {KDT}
\bibfield{author}{\bibinfo{person}{Jon~Louis Bentley}.}
  \bibinfo{year}{1975}\natexlab{}.
\newblock \showarticletitle{Multidimensional Binary Search Trees Used for
  Associative Searching}.
\newblock  (\bibinfo{year}{1975}), \bibinfo{pages}{509--517}.
\newblock


\bibitem[\protect\citeauthoryear{Bonnabel}{Bonnabel}{2013}]%
        {RSGD}
\bibfield{author}{\bibinfo{person}{Silvere Bonnabel}.}
  \bibinfo{year}{2013}\natexlab{}.
\newblock \showarticletitle{Stochastic Gradient Descent on Riemannian
  Manifolds}.
\newblock \bibinfo{journal}{\emph{IEEE Trans. Automat. Control}}
  \bibinfo{volume}{58}, \bibinfo{number}{9} (\bibinfo{year}{2013}),
  \bibinfo{pages}{2217--2229}.
\newblock


\bibitem[\protect\citeauthoryear{Chami, Ying, Ré, and Leskovec}{Chami
  et~al\mbox{.}}{2019}]%
        {HGCN}
\bibfield{author}{\bibinfo{person}{Ines Chami}, \bibinfo{person}{Zhitao Ying},
  \bibinfo{person}{Christopher Ré}, {and} \bibinfo{person}{Jure Leskovec}.}
  \bibinfo{year}{2019}\natexlab{}.
\newblock \showarticletitle{Hyperbolic Graph Convolutional Neural Networks}. In
  \bibinfo{booktitle}{\emph{NeurIPS}}. \bibinfo{pages}{4869--4880}.
\newblock


\bibitem[\protect\citeauthoryear{Chen, Wu, Hong, Zhang, and Wang}{Chen
  et~al\mbox{.}}{2020}]%
        {LRGCCF}
\bibfield{author}{\bibinfo{person}{Lei Chen}, \bibinfo{person}{Le Wu},
  \bibinfo{person}{Richang Hong}, \bibinfo{person}{Kun Zhang}, {and}
  \bibinfo{person}{Meng Wang}.} \bibinfo{year}{2020}\natexlab{}.
\newblock \showarticletitle{Revisiting Graph Based Collaborative Filtering: A
  Linear Residual Graph Convolutional Network Approach}. In
  \bibinfo{booktitle}{\emph{AAAI}}. \bibinfo{pages}{27--34}.
\newblock


\bibitem[\protect\citeauthoryear{Ebesu, Shen, and Fang}{Ebesu
  et~al\mbox{.}}{2018}]%
        {CMN}
\bibfield{author}{\bibinfo{person}{Travis Ebesu}, \bibinfo{person}{Bin Shen},
  {and} \bibinfo{person}{Yi Fang}.} \bibinfo{year}{2018}\natexlab{}.
\newblock \showarticletitle{Collaborative Memory Network for Recommendation
  Systems}. In \bibinfo{booktitle}{\emph{SIGIR}}. \bibinfo{pages}{515--524}.
\newblock


\bibitem[\protect\citeauthoryear{Ganea, Bécigneul, and Hofmann}{Ganea
  et~al\mbox{.}}{2018a}]%
        {hyperrelated2}
\bibfield{author}{\bibinfo{person}{Octavian-Eugen Ganea}, \bibinfo{person}{Gary
  Bécigneul}, {and} \bibinfo{person}{Thomas Hofmann}.}
  \bibinfo{year}{2018}\natexlab{a}.
\newblock \showarticletitle{Hyperbolic Entailment Cones for Learning
  Hierarchical Embeddings}. In \bibinfo{booktitle}{\emph{ICML}}.
  \bibinfo{pages}{1632--1641}.
\newblock


\bibitem[\protect\citeauthoryear{Ganea, Bécigneul, and Hofmann}{Ganea
  et~al\mbox{.}}{2018b}]%
        {HNN}
\bibfield{author}{\bibinfo{person}{Octavian-Eugen Ganea}, \bibinfo{person}{Gary
  Bécigneul}, {and} \bibinfo{person}{Thomas Hofmann}.}
  \bibinfo{year}{2018}\natexlab{b}.
\newblock \showarticletitle{Hyperbolic Neural Networks}. In
  \bibinfo{booktitle}{\emph{NeurIPS}}. \bibinfo{pages}{5350--5360}.
\newblock


\bibitem[\protect\citeauthoryear{Guillaume and Latapy}{Guillaume and
  Latapy}{2006}]%
        {BipaUI}
\bibfield{author}{\bibinfo{person}{Jean-Loup Guillaume} {and}
  \bibinfo{person}{Matthieu Latapy}.} \bibinfo{year}{2006}\natexlab{}.
\newblock \showarticletitle{Bipartite graphs as models of complex networks}.
\newblock  (\bibinfo{year}{2006}), \bibinfo{pages}{795–813}.
\newblock


\bibitem[\protect\citeauthoryear{He and Chua}{He and Chua}{2017}]%
        {NFM}
\bibfield{author}{\bibinfo{person}{Xiangnan He} {and} \bibinfo{person}{Tat-Seng
  Chua}.} \bibinfo{year}{2017}\natexlab{}.
\newblock \showarticletitle{Neural Factorization Machines for Sparse Predictive
  Analytics}. In \bibinfo{booktitle}{\emph{SIGIR}}. \bibinfo{pages}{355--364}.
\newblock


\bibitem[\protect\citeauthoryear{He, Deng, Wang, Li, Zhang, and Wang}{He
  et~al\mbox{.}}{2020}]%
        {LightGCN}
\bibfield{author}{\bibinfo{person}{Xiangnan He}, \bibinfo{person}{Kuan Deng},
  \bibinfo{person}{Xiang Wang}, \bibinfo{person}{Yan Li},
  \bibinfo{person}{Yong-Dong Zhang}, {and} \bibinfo{person}{Meng Wang}.}
  \bibinfo{year}{2020}\natexlab{}.
\newblock \showarticletitle{LightGCN: Simplifying and Powering Graph
  Convolution Network for Recommendation}. In
  \bibinfo{booktitle}{\emph{SIGIR}}. \bibinfo{pages}{639--648}.
\newblock


\bibitem[\protect\citeauthoryear{He, Liao, Zhang, Nie, and Xia~Hu}{He
  et~al\mbox{.}}{2017}]%
        {NCF}
\bibfield{author}{\bibinfo{person}{Xiangnan He}, \bibinfo{person}{Lizi Liao},
  \bibinfo{person}{Hanwang Zhang}, \bibinfo{person}{Liqiang Nie}, {and}
  \bibinfo{person}{Tat-Seng~Chua Xia~Hu}.} \bibinfo{year}{2017}\natexlab{}.
\newblock \showarticletitle{Neural collaborative filtering}. In
  \bibinfo{booktitle}{\emph{WWW}}. \bibinfo{pages}{173--182}.
\newblock


\bibitem[\protect\citeauthoryear{He, Zhang, Kan, and Chua}{He
  et~al\mbox{.}}{2016}]%
        {CF3}
\bibfield{author}{\bibinfo{person}{Xiangnan He}, \bibinfo{person}{Hanwang
  Zhang}, \bibinfo{person}{Min-Yen Kan}, {and} \bibinfo{person}{Tat-Seng
  Chua}.} \bibinfo{year}{2016}\natexlab{}.
\newblock \showarticletitle{Fast Matrix Factorization for Online Recommendation
  with Implicit Feedback}. In \bibinfo{booktitle}{\emph{SIGIR}}.
  \bibinfo{pages}{549--558}.
\newblock


\bibitem[\protect\citeauthoryear{Hu, Koren, and Volinsky}{Hu
  et~al\mbox{.}}{2008}]%
        {CF4}
\bibfield{author}{\bibinfo{person}{Yifan Hu}, \bibinfo{person}{Yehuda Koren},
  {and} \bibinfo{person}{Chris Volinsky}.} \bibinfo{year}{2008}\natexlab{}.
\newblock \showarticletitle{Collaborative Filtering for Implicit Feedback
  Datasets}. In \bibinfo{booktitle}{\emph{ICDM}}. \bibinfo{pages}{263--272}.
\newblock


\bibitem[\protect\citeauthoryear{Jian~Tang}{Jian~Tang}{2015}]%
        {LINE}
\bibfield{author}{\bibinfo{person}{Mingzhe Wang Ming Zhang Jun Yan Qiaozhu~Mei
  Jian~Tang, Meng~Qu}.} \bibinfo{year}{2015}\natexlab{}.
\newblock \showarticletitle{LINE: Large-scale Information Network Embedding}.
  In \bibinfo{booktitle}{\emph{WWW}}. \bibinfo{pages}{1067--1077}.
\newblock


\bibitem[\protect\citeauthoryear{Jiang, Hu, Fang, and Shi}{Jiang
  et~al\mbox{.}}{2020}]%
        {MMCF}
\bibfield{author}{\bibinfo{person}{Xunqiang Jiang}, \bibinfo{person}{Binbin
  Hu}, \bibinfo{person}{Yuan Fang}, {and} \bibinfo{person}{Chuan Shi}.}
  \bibinfo{year}{2020}\natexlab{}.
\newblock \showarticletitle{Multiplex Memory Network for Collaborative
  Filtering}. In \bibinfo{booktitle}{\emph{SDM}}. \bibinfo{pages}{91--99}.
\newblock


\bibitem[\protect\citeauthoryear{Koren}{Koren}{2008}]%
        {SVD}
\bibfield{author}{\bibinfo{person}{Yehuda Koren}.}
  \bibinfo{year}{2008}\natexlab{}.
\newblock \showarticletitle{Factorization meets the neighborhood: a
  multifaceted collaborative filtering model}. In
  \bibinfo{booktitle}{\emph{Proceedings of the 14th ACM SIGKDD International
  Conference on Knowledge Discovery and Data Mining}}.
  \bibinfo{pages}{426--434}.
\newblock


\bibitem[\protect\citeauthoryear{Koren, Bell, and Volinsky}{Koren
  et~al\mbox{.}}{2009}]%
        {MF}
\bibfield{author}{\bibinfo{person}{Yehuda Koren}, \bibinfo{person}{Robert~M.
  Bell}, {and} \bibinfo{person}{Chris Volinsky}.}
  \bibinfo{year}{2009}\natexlab{}.
\newblock \showarticletitle{Matrix Factorization Techniques for Recommender
  Systems}.
\newblock \bibinfo{journal}{\emph{IEEE Computer}} \bibinfo{volume}{42},
  \bibinfo{number}{8} (\bibinfo{date}{August} \bibinfo{year}{2009}),
  \bibinfo{pages}{30--37}.
\newblock


\bibitem[\protect\citeauthoryear{Krioukov, Papadopoulos, Kitsak, Vahdat, and
  Boguñá}{Krioukov et~al\mbox{.}}{2010}]%
        {hp1}
\bibfield{author}{\bibinfo{person}{Dmitri~V. Krioukov},
  \bibinfo{person}{Fragkiskos Papadopoulos}, \bibinfo{person}{Maksim Kitsak},
  \bibinfo{person}{Amin Vahdat}, {and} \bibinfo{person}{Marián Boguñá}.}
  \bibinfo{year}{2010}\natexlab{}.
\newblock \showarticletitle{Hyperbolic Geometry of Complex Networks}. In
  \bibinfo{booktitle}{\emph{CoRR abs/1006.5169}}.
\newblock


\bibitem[\protect\citeauthoryear{Linden, Smith, and York}{Linden
  et~al\mbox{.}}{2003}]%
        {item_item2}
\bibfield{author}{\bibinfo{person}{Greg Linden}, \bibinfo{person}{Brent Smith},
  {and} \bibinfo{person}{Jeremy York}.} \bibinfo{year}{2003}\natexlab{}.
\newblock \showarticletitle{Amazon.com Recommendations: Item-to-Item
  Collaborative Filtering}. In \bibinfo{booktitle}{\emph{IEEE Internet
  Computing}}. \bibinfo{pages}{76--80}.
\newblock


\bibitem[\protect\citeauthoryear{Liu, Moore, and Gray}{Liu
  et~al\mbox{.}}{2006}]%
        {BT2}
\bibfield{author}{\bibinfo{person}{Ting Liu}, \bibinfo{person}{Andrew~W.
  Moore}, {and} \bibinfo{person}{Alexander~G. Gray}.}
  \bibinfo{year}{2006}\natexlab{}.
\newblock \showarticletitle{New Algorithms for Efficient High-Dimensional
  Nonparametric Classification}.
\newblock  (\bibinfo{year}{2006}), \bibinfo{pages}{1135--1158}.
\newblock


\bibitem[\protect\citeauthoryear{Maas, L., Hannun, and Ng}{Maas
  et~al\mbox{.}}{2013}]%
        {LeakyReLU}
\bibfield{author}{\bibinfo{person}{Maas}, \bibinfo{person}{Andrew L.},
  \bibinfo{person}{Awni~Y. Hannun}, {and} \bibinfo{person}{Andrew~Y. Ng}.}
  \bibinfo{year}{2013}\natexlab{}.
\newblock \showarticletitle{Rectifier nonlinearities improve neural network
  acoustic models}. In \bibinfo{booktitle}{\emph{ICML}}.
\newblock


\bibitem[\protect\citeauthoryear{Nickel and Kiela}{Nickel and Kiela}{2017}]%
        {Poincare}
\bibfield{author}{\bibinfo{person}{Maximilian Nickel} {and}
  \bibinfo{person}{Douwe Kiela}.} \bibinfo{year}{2017}\natexlab{}.
\newblock \showarticletitle{Poincare Embeddings for Learning Hierarchical
  Representations}. In \bibinfo{booktitle}{\emph{NeurIPS}}.
  \bibinfo{pages}{6338--6347}.
\newblock


\bibitem[\protect\citeauthoryear{Nickel and Kiela}{Nickel and Kiela}{2018}]%
        {hyperrelated1}
\bibfield{author}{\bibinfo{person}{Maximilian Nickel} {and}
  \bibinfo{person}{Douwe Kiela}.} \bibinfo{year}{2018}\natexlab{}.
\newblock \showarticletitle{Learning Continuous Hierarchies in the Lorentz
  Model of Hyperbolic Geometry}. In \bibinfo{booktitle}{\emph{ICML}}.
  \bibinfo{pages}{3776--3785}.
\newblock


\bibitem[\protect\citeauthoryear{Omohundro}{Omohundro}{1989}]%
        {BT1}
\bibfield{author}{\bibinfo{person}{Stephen~M Omohundro}.}
  \bibinfo{year}{1989}\natexlab{}.
\newblock \bibinfo{booktitle}{\emph{Five balltree construction algorithms}}.
\newblock


\bibitem[\protect\citeauthoryear{Pei, Wei, Chang, Lei, and Yang}{Pei
  et~al\mbox{.}}{2020}]%
        {Gemo}
\bibfield{author}{\bibinfo{person}{Hongbin Pei}, \bibinfo{person}{Bingzhe Wei},
  \bibinfo{person}{Kevin Chen-Chuan Chang}, \bibinfo{person}{Yu Lei}, {and}
  \bibinfo{person}{Bo Yang}.} \bibinfo{year}{2020}\natexlab{}.
\newblock \showarticletitle{Geom-GCN: Geometric Graph Convolutional Networks}.
  In \bibinfo{booktitle}{\emph{ICLR}}.
\newblock


\bibitem[\protect\citeauthoryear{Perozzi, Al-Rfou, and Skiena}{Perozzi
  et~al\mbox{.}}{2014}]%
        {DeepWalk}
\bibfield{author}{\bibinfo{person}{Bryan Perozzi}, \bibinfo{person}{Rami
  Al-Rfou}, {and} \bibinfo{person}{Steven Skiena}.}
  \bibinfo{year}{2014}\natexlab{}.
\newblock \showarticletitle{DeepWalk: online learning of social
  representations}. In \bibinfo{booktitle}{\emph{KDD}}.
  \bibinfo{pages}{701--710}.
\newblock


\bibitem[\protect\citeauthoryear{Qu, Bai, Zhang, Nie, and Tang}{Qu
  et~al\mbox{.}}{2019}]%
        {KNI}
\bibfield{author}{\bibinfo{person}{Yanru Qu}, \bibinfo{person}{Ting Bai},
  \bibinfo{person}{Weinan Zhang}, \bibinfo{person}{Jian-Yun Nie}, {and}
  \bibinfo{person}{Jian Tang}.} \bibinfo{year}{2019}\natexlab{}.
\newblock \showarticletitle{An End-to-End Neighborhood-based Interaction Model
  for Knowledge-enhanced Recommendation}. In \bibinfo{booktitle}{\emph{CoRR
  abs/1908.04032}}.
\newblock


\bibitem[\protect\citeauthoryear{Ravasz and Barabasi}{Ravasz and
  Barabasi}{2003}]%
        {HOCN}
\bibfield{author}{\bibinfo{person}{Erzsebet Ravasz} {and}
  \bibinfo{person}{Albert-Laszlo Barabasi}.} \bibinfo{year}{2003}\natexlab{}.
\newblock \showarticletitle{Hierarchical organization in complex networks}.
\newblock \bibinfo{journal}{\emph{Physical review E}} \bibinfo{volume}{67},
  \bibinfo{number}{2} (\bibinfo{year}{2003}), \bibinfo{pages}{026112}.
\newblock


\bibitem[\protect\citeauthoryear{Rendle}{Rendle}{2010}]%
        {FM}
\bibfield{author}{\bibinfo{person}{Steffen Rendle}.}
  \bibinfo{year}{2010}\natexlab{}.
\newblock \showarticletitle{Factorization Machines}. In
  \bibinfo{booktitle}{\emph{ICDM}}. \bibinfo{pages}{995--1000}.
\newblock


\bibitem[\protect\citeauthoryear{Ribeiro, Saverese, and Figueiredo}{Ribeiro
  et~al\mbox{.}}{2017}]%
        {struc2vec}
\bibfield{author}{\bibinfo{person}{Leonardo Filipe~Rodrigues Ribeiro},
  \bibinfo{person}{Pedro H.~P. Saverese}, {and} \bibinfo{person}{Daniel~R.
  Figueiredo}.} \bibinfo{year}{2017}\natexlab{}.
\newblock \showarticletitle{struc2vec: Learning Node Representations from
  Structural Identity}. In \bibinfo{booktitle}{\emph{KDD}}.
  \bibinfo{pages}{385--394}.
\newblock


\bibitem[\protect\citeauthoryear{Ricci, Rokach, and Shapira}{Ricci
  et~al\mbox{.}}{2011}]%
        {RecBook}
\bibfield{author}{\bibinfo{person}{Francesco Ricci}, \bibinfo{person}{Lior
  Rokach}, {and} \bibinfo{person}{Bracha Shapira}.}
  \bibinfo{year}{2011}\natexlab{}.
\newblock \showarticletitle{Introduction to recommender systems handbook}.
\newblock  (\bibinfo{year}{2011}).
\newblock


\bibitem[\protect\citeauthoryear{Sala, Sa, Gu, and Ré}{Sala
  et~al\mbox{.}}{2018}]%
        {hyperrelated3}
\bibfield{author}{\bibinfo{person}{Frederic Sala},
  \bibinfo{person}{Christopher~De Sa}, \bibinfo{person}{Albert Gu}, {and}
  \bibinfo{person}{Christopher Ré}.} \bibinfo{year}{2018}\natexlab{}.
\newblock \showarticletitle{Representation Tradeoffs for Hyperbolic
  Embeddings}. In \bibinfo{booktitle}{\emph{ICML}}.
  \bibinfo{pages}{4457--4466}.
\newblock


\bibitem[\protect\citeauthoryear{Sarwar, Karypis, Konstan, and Riedl}{Sarwar
  et~al\mbox{.}}{2001}]%
        {CF1}
\bibfield{author}{\bibinfo{person}{Badrul~Munir Sarwar},
  \bibinfo{person}{George Karypis}, \bibinfo{person}{Joseph~A. Konstan}, {and}
  \bibinfo{person}{John Riedl}.} \bibinfo{year}{2001}\natexlab{}.
\newblock \showarticletitle{Item-based collaborative filtering recommendation
  algorithms}. In \bibinfo{booktitle}{\emph{WWW}}. \bibinfo{pages}{285--295}.
\newblock


\bibitem[\protect\citeauthoryear{Sun, Zhang, Ma, Coates, Guo, Tang, and He}{Sun
  et~al\mbox{.}}{2019}]%
        {GCCF}
\bibfield{author}{\bibinfo{person}{Jianing Sun}, \bibinfo{person}{Yingxue
  Zhang}, \bibinfo{person}{Chen Ma}, \bibinfo{person}{Mark Coates},
  \bibinfo{person}{Huifeng Guo}, \bibinfo{person}{Ruiming Tang}, {and}
  \bibinfo{person}{Xiuqiang He}.} \bibinfo{year}{2019}\natexlab{}.
\newblock \showarticletitle{Multi-graph Convolution Collaborative Filtering}.
  In \bibinfo{booktitle}{\emph{ICDM}}. \bibinfo{pages}{1306--1311}.
\newblock


\bibitem[\protect\citeauthoryear{Tifrea, Bécigneul, and Ganea}{Tifrea
  et~al\mbox{.}}{2019}]%
        {PoincareGlove}
\bibfield{author}{\bibinfo{person}{Alexandru Tifrea}, \bibinfo{person}{Gary
  Bécigneul}, {and} \bibinfo{person}{Octavian-Eugen Ganea}.}
  \bibinfo{year}{2019}\natexlab{}.
\newblock \showarticletitle{Poincaré Glove: Hyperbolic Word Embeddings}. In
  \bibinfo{booktitle}{\emph{ICLR}}.
\newblock


\bibitem[\protect\citeauthoryear{Tran, Tay, Zhang, Cong, and Li}{Tran
  et~al\mbox{.}}{2020}]%
        {HyperML}
\bibfield{author}{\bibinfo{person}{Lucas~Vinh Tran}, \bibinfo{person}{Yi Tay},
  \bibinfo{person}{Shuai Zhang}, \bibinfo{person}{Gao Cong}, {and}
  \bibinfo{person}{Xiaoli Li}.} \bibinfo{year}{2020}\natexlab{}.
\newblock \showarticletitle{HyperML: A Boosting Metric Learning Approach in
  Hyperbolic Space for Recommender Systems}. In
  \bibinfo{booktitle}{\emph{WSDM}}. \bibinfo{pages}{609--617}.
\newblock


\bibitem[\protect\citeauthoryear{Ungar}{Ungar}{2001}]%
        {gyrovector1}
\bibfield{author}{\bibinfo{person}{Abraham~Albert Ungar}.}
  \bibinfo{year}{2001}\natexlab{}.
\newblock \showarticletitle{Hyperbolic trigonometry and its application in the
  Poincare ball model of hyperbolic geometry}.
\newblock  (\bibinfo{year}{2001}), \bibinfo{pages}{135--147}.
\newblock


\bibitem[\protect\citeauthoryear{Ungar}{Ungar}{2008}]%
        {gyrovector2}
\bibfield{author}{\bibinfo{person}{Abraham~Albert Ungar}.}
  \bibinfo{year}{2008}\natexlab{}.
\newblock \showarticletitle{A gyrovector space approach to hyperbolic
  geometry}.
\newblock  (\bibinfo{year}{2008}), \bibinfo{pages}{1--194}.
\newblock


\bibitem[\protect\citeauthoryear{Vinh, Tay, Zhang, Cong, and Li}{Vinh
  et~al\mbox{.}}{2018}]%
        {HyperBPR}
\bibfield{author}{\bibinfo{person}{Tran Dang~Quang Vinh}, \bibinfo{person}{Yi
  Tay}, \bibinfo{person}{Shuai Zhang}, \bibinfo{person}{Gao Cong}, {and}
  \bibinfo{person}{Xiao-Li Li}.} \bibinfo{year}{2018}\natexlab{}.
\newblock \showarticletitle{AAAI}. In \bibinfo{booktitle}{\emph{CoRR
  abs/1809.01703}}.
\newblock


\bibitem[\protect\citeauthoryear{Wang, Cui, and Zhu}{Wang
  et~al\mbox{.}}{2016}]%
        {SDNE}
\bibfield{author}{\bibinfo{person}{Daixin Wang}, \bibinfo{person}{Peng Cui},
  {and} \bibinfo{person}{Wenwu Zhu}.} \bibinfo{year}{2016}\natexlab{}.
\newblock \showarticletitle{Structural Deep Network Embedding}. In
  \bibinfo{booktitle}{\emph{KDD}}. \bibinfo{pages}{1225--1234}.
\newblock


\bibitem[\protect\citeauthoryear{Wang, de~Vries, and Reinders}{Wang
  et~al\mbox{.}}{2006}]%
        {item_item}
\bibfield{author}{\bibinfo{person}{Jun Wang}, \bibinfo{person}{Arjen~P. de
  Vries}, {and} \bibinfo{person}{Marcel J.~T. Reinders}.}
  \bibinfo{year}{2006}\natexlab{}.
\newblock \showarticletitle{Unifying user-based and item-based collaborative
  filtering approaches by similarity fusion}. In
  \bibinfo{booktitle}{\emph{SIGIR}}. \bibinfo{pages}{501--508}.
\newblock


\bibitem[\protect\citeauthoryear{Wang, He, Wang, Feng, and Chua}{Wang
  et~al\mbox{.}}{2019a}]%
        {NGCF}
\bibfield{author}{\bibinfo{person}{Xiang Wang}, \bibinfo{person}{Xiangnan He},
  \bibinfo{person}{Meng Wang}, \bibinfo{person}{Fuli Feng}, {and}
  \bibinfo{person}{Tat-Seng Chua}.} \bibinfo{year}{2019}\natexlab{a}.
\newblock \showarticletitle{Neural Graph Collaborative Filtering}. In
  \bibinfo{booktitle}{\emph{SIGIR}}. \bibinfo{pages}{165--174}.
\newblock


\bibitem[\protect\citeauthoryear{Wang, Zhang, and Shi}{Wang
  et~al\mbox{.}}{2019b}]%
        {HHNE}
\bibfield{author}{\bibinfo{person}{Xiao Wang}, \bibinfo{person}{Yiding Zhang},
  {and} \bibinfo{person}{Chuan Shi}.} \bibinfo{year}{2019}\natexlab{b}.
\newblock \showarticletitle{Hyperbolic Heterogeneous Information Network
  Embedding}. In \bibinfo{booktitle}{\emph{AAAI}}. \bibinfo{pages}{5337--5344}.
\newblock


\bibitem[\protect\citeauthoryear{Wu, Sun, Fu, Hong, Wang, and Wang}{Wu
  et~al\mbox{.}}{2019}]%
        {Diffnet}
\bibfield{author}{\bibinfo{person}{Le Wu}, \bibinfo{person}{Peijie Sun},
  \bibinfo{person}{Yanjie Fu}, \bibinfo{person}{Richang Hong},
  \bibinfo{person}{Xiting Wang}, {and} \bibinfo{person}{Meng Wang}.}
  \bibinfo{year}{2019}\natexlab{}.
\newblock \showarticletitle{A Neural Influence Diffusion Model for Social
  Recommendation}. In \bibinfo{booktitle}{\emph{SIGIR}}.
  \bibinfo{pages}{235--244}.
\newblock


\bibitem[\protect\citeauthoryear{Wu, DuBois, Zheng, and Ester}{Wu
  et~al\mbox{.}}{2016}]%
        {CDAE}
\bibfield{author}{\bibinfo{person}{Yao Wu}, \bibinfo{person}{Christopher
  DuBois}, \bibinfo{person}{Alice~X. Zheng}, {and} \bibinfo{person}{Martin
  Ester}.} \bibinfo{year}{2016}\natexlab{}.
\newblock \showarticletitle{Collaborative Denoising Auto-Encoders for Top-N
  Recommender Systems}. In \bibinfo{booktitle}{\emph{WSDM}}.
  \bibinfo{pages}{153--162}.
\newblock


\bibitem[\protect\citeauthoryear{Xue, Dai, Zhang, Huang, and Chen}{Xue
  et~al\mbox{.}}{2017}]%
        {DMF}
\bibfield{author}{\bibinfo{person}{Hong-Jian Xue}, \bibinfo{person}{Xinyu Dai},
  \bibinfo{person}{Jianbing Zhang}, \bibinfo{person}{Shujian Huang}, {and}
  \bibinfo{person}{Jiajun Chen}.} \bibinfo{year}{2017}\natexlab{}.
\newblock \showarticletitle{Deep Matrix Factorization Models for Recommender
  Systems}. In \bibinfo{booktitle}{\emph{IJCAI}}. \bibinfo{pages}{3203--3209}.
\newblock


\bibitem[\protect\citeauthoryear{Zhang, Shen, Liu, He, Luan, and Chua}{Zhang
  et~al\mbox{.}}{2016}]%
        {CF2}
\bibfield{author}{\bibinfo{person}{Hanwang Zhang}, \bibinfo{person}{Fumin
  Shen}, \bibinfo{person}{Wei Liu}, \bibinfo{person}{Xiangnan He},
  \bibinfo{person}{Huanbo Luan}, {and} \bibinfo{person}{Tat-Seng Chua}.}
  \bibinfo{year}{2016}\natexlab{}.
\newblock \showarticletitle{Discrete Collaborative Filtering}. In
  \bibinfo{booktitle}{\emph{SIGIR}}. \bibinfo{pages}{325--334}.
\newblock


\bibitem[\protect\citeauthoryear{Zhang, Wang, Jiang, Shi, and Ye}{Zhang
  et~al\mbox{.}}{2019}]%
        {HAT}
\bibfield{author}{\bibinfo{person}{Yiding Zhang}, \bibinfo{person}{Xiao Wang},
  \bibinfo{person}{Xunqiang Jiang}, \bibinfo{person}{Chuan Shi}, {and}
  \bibinfo{person}{Yanfang Ye}.} \bibinfo{year}{2019}\natexlab{}.
\newblock \showarticletitle{Hyperbolic Graph Attention Network}. In
  \bibinfo{booktitle}{\emph{CoRR abs/1912.03046}}.
\newblock


\bibitem[\protect\citeauthoryear{Zheng, Lu, Jiang, Zhang, and Yu}{Zheng
  et~al\mbox{.}}{2018}]%
        {SpectralCF}
\bibfield{author}{\bibinfo{person}{Lei Zheng}, \bibinfo{person}{Chun-Ta Lu},
  \bibinfo{person}{Fei Jiang}, \bibinfo{person}{Jiawei Zhang}, {and}
  \bibinfo{person}{Philip~S. Yu}.} \bibinfo{year}{2018}\natexlab{}.
\newblock \showarticletitle{Spectral collaborative filtering}. In
  \bibinfo{booktitle}{\emph{RecSys}}. \bibinfo{pages}{311--319}.
\newblock


\bibitem[\protect\citeauthoryear{Çaglar Gülçehre, Denil, Malinowski, Razavi,
  Pascanu, Hermann, Battaglia, Bapst, Raposo, Santoro, and de~Freitas}{Çaglar
  Gülçehre et~al\mbox{.}}{2019}]%
        {hp2}
\bibfield{author}{\bibinfo{person}{Çaglar Gülçehre}, \bibinfo{person}{Misha
  Denil}, \bibinfo{person}{Mateusz Malinowski}, \bibinfo{person}{Ali Razavi},
  \bibinfo{person}{Razvan Pascanu}, \bibinfo{person}{Karl~Moritz Hermann},
  \bibinfo{person}{Peter~W. Battaglia}, \bibinfo{person}{Victor Bapst},
  \bibinfo{person}{David Raposo}, \bibinfo{person}{Adam Santoro}, {and}
  \bibinfo{person}{Nando de Freitas}.} \bibinfo{year}{2019}\natexlab{}.
\newblock \showarticletitle{Hyperbolic Attention Networks}. In
  \bibinfo{booktitle}{\emph{ICLR}}.
\newblock


\end{thebibliography}

\appendix

\end{document}